\documentclass[sigconf]{acmart}

\usepackage{kotex}       
\usepackage{times}       

\usepackage{mathrsfs}

\usepackage{graphicx}
\usepackage{xcolor}
\usepackage{subcaption}

\usepackage{booktabs}    
\usepackage{multirow}

\usepackage{algorithm}
\usepackage{algorithmicx}     
\usepackage{algpseudocode}    

\usepackage[title]{appendix}
\usepackage{chngcntr}
\usepackage{textcomp}
\usepackage{manyfoot}
\usepackage{listings}
\usepackage{soul}
\usepackage{url}
\usepackage[switch]{lineno}

\usepackage{graphicx}%
\usepackage{multirow}%
\usepackage{amsmath,amsfonts}%
\usepackage{amsthm}%
\usepackage{mathrsfs}%
\usepackage[title]{appendix}%
\usepackage{xcolor}%
\usepackage{textcomp}%
\usepackage{manyfoot}%
\usepackage{booktabs}%
\usepackage{algpseudocode}%
\usepackage{listings}%
\usepackage{afterpage}

\setcopyright{none}
\copyrightyear{}
\acmYear{}
\acmDOI{}

\acmConference[]{}{}{}

\acmISBN{}

\begin{document}

\title[Personalized and Multi-View Representation for Federated Cold-Start Recommendation]{Personalized and Multi-View Representation for Federated Cold-Start Recommendation}

\author{Jaehyung Lim}
\affiliation{%
  \institution{Pohang University of\\ Science and Technology}
  \city{Pohang}
  \country{Republic of Korea}}
\email{jaehyunglim@postech.ac.kr}

\author{Wonbin Kweon}
\affiliation{%
  \institution{University of Illinois  Urbana-Champaign}
  \city{Champaign}
  \state{IL}
  \country{USA}}
\email{wonbin@illinois.edu}

\author{Woojoo Kim}
\affiliation{%
  \institution{Pohang University of\\ Science and Technology}
  \city{Pohang}
  \country{Republic of Korea}}
\email{kimuj0103@postech.ac.kr}

\author{Junyoung Kim}
\affiliation{%
  \institution{Pohang University of\\ Science and Technology}
  \city{Pohang}
  \country{Republic of Korea}}
\email{junyoungkim@postech.ac.kr}

\author{Dongha Kim}
\affiliation{%
  \institution{Pohang University of\\ Science and Technology}
  \city{Pohang}
  \country{Republic of Korea}}
\email{dhkim0317@postech.ac.kr}

\author{Hwanjo Yu}
\authornote{Corresponding author}
\affiliation{%
  \institution{Pohang University of\\ Science and Technology}
  \city{Pohang}
  \country{Republic of Korea}}
\email{hwanjoyu@postech.ac.kr}

\renewcommand{\shortauthors}{Jaehyung Lim et al.}

\begin{abstract}
Federated recommendation (FedRec) enables personalized modeling without centralizing users’ interaction histories, but most existing methods assume a fixed item pool and thus overlook the practical cold-item setting where new items continuously arrive. Under the dual-sided constraint, where the server cannot access clients’ interactions while clients cannot access the server’s proprietary item attribute features, prior federated cold-start recommendation approaches suffer from three structural limitations: a lack of personalization, compositionality failure caused by forcing heterogeneous semantics into a single embedding space, and training- and communication-inefficiency arising from explicit alignment between separate collaborative and attribute representations. To address these challenges, we propose  \underline{\textbf{P}}ersonalized and \underline{\textbf{M}}ulti-view Representation for \underline{\textbf{F}}ederated Cold-Start \underline{\textbf{Rec}}ommendation (\textbf{PMFRec}). PMFRec learns a personalized representation generator to produce user-specific item representations from attribute features, and introduces a global multi-view encoder with item-adaptive gating and an orthogonality objective to capture complementary semantic views while reducing cross-view redundancy. In addition, PMFRec fuses collaborative and attribute knowledge into a single exchanged item representation, eliminating the need for an explicit client-side regularizer and reducing communication overhead. Extensive experiments on real-world datasets show that PMFRec consistently outperforms strong baselines in cold-item recommendation and further improves user-level fairness, warm-scenario adaptability, and robustness under Local Differential Privacy (LDP).
\end{abstract}



\received{20 February 2007}
\received[revised]{12 March 2009}
\received[accepted]{5 June 2009}

\maketitle

\section{Introduction}
\label{sec:introduction}

Recommender systems are increasingly crucial as online platforms scale in both users and items \citep{ncf, lightgcn}. At the same time, privacy regulations (e.g., CCPA \citep{ccpa}, GDPR \citep{gdpr}) make centralized training—collecting users’ interaction histories on a server—progressively less feasible. To address these limitations, Federated Learning (FL) \citep{first_fl_paper} has been introduced, and its adaptation to recommender systems has led to Federated Recommendation (FedRec) \citep{fcf, fedrap, fediar, fedcia}. FedRec addresses this by keeping user data on-device and training models in a distributed manner, enabling personalization with stronger privacy protection.

Most prior FedRec studies assume a fixed item pool, whereas real-world services continuously introduce new items, making cold-item recommendation to existing users a practical necessity. This is a key instance of the cold-start problem in recommender systems \citep{rs_survey, hybrid_2, heater, cold_cf_2}.
Despite its importance, federated cold-item recommendation remains under-explored. Early work \citep{first_fedcold} formulates the problem in federated settings, but it neither addresses the practically common case of \emph{zero-interaction} new items nor exploits item attribute features that are crucial for cold-start \citep{rs_survey, hybrid_2, heater, cold_cf_2}.
More recently, \citep{ifedrec} targets zero-interaction cold items and introduces the \emph{dual-sided constraint} between the server (holding attribute features) and clients (holding interactions).
It learns a single global attribute-feature-to-embedding mapping from aggregated client signals, enabling the server to generate representations for zero-interaction new items from their attribute features. While effective, this single global attribute-feature-to-embedding mapping paradigm faces three structural limitations in federated cold-start recommendation.
First, it suffers from a lack of personalization. Second, it encounters compositionality failure. Third, it is training- and communication-inefficient. 

\begin{table}[t]
\centering
\caption{Toy example illustrating paper recommendation: each item corresponds to a paper, and the category indicates its research area.}
\label{tab:toy_example}

\begin{tabular}{l|ccccc}
\toprule
Category & Drug & ML & Graph & Chemistry & Recommendation \\
\midrule
Item A & \checkmark & \checkmark & \checkmark &  &  \\
Item B & \checkmark &  &  & \checkmark &  \\
Item C &  & \checkmark & \checkmark &  & \checkmark \\
\bottomrule
\end{tabular}
\end{table}

First, the lack of personalization arises from two compounding factors:
(i) prior methods rely on a unified server-side mapping that generates item representations in a user-agnostic manner, and
(ii) under the dual-sided constraint, clients cannot access raw item attribute features, which fundamentally prevents them from learning personalized attribute-to-embedding mappings locally.
As a result, cold-item representations are produced by a shared global function that captures average patterns across users rather than individualized preference structures.
This limitation becomes particularly severe in cold-item scenarios, where no interaction history is available and representations must therefore be generated solely from attribute features.

To illustrate, consider the toy example in Table~\ref{tab:toy_example}, where each item corresponds to a newly introduced paper and each attribute dimension indicates a research category.
Suppose item $A$ is related to \emph{Drug}, \emph{ML}, and \emph{Graph}, item $B$ is related to \emph{Drug} and \emph{Chemistry}, and item $C$ is related to \emph{ML}, \emph{Graph}, and \emph{Recommendation}.
Now consider two users with different interests:
one user mainly prefers drug-related studies, whereas another is more interested in machine learning or graph-oriented papers.
Under an ideal personalized generator, the first user should regard item $A$ as being closer to item $B$, while the second user should regard item $A$ as being closer to item $C$.
However, existing personalization-agnostic approaches cannot express such user-specific differences.
Because they rely on a single shared mapping, the same attribute configuration is translated into essentially the same cold-item representation for all users,
thereby diluting the individual preference signals that are most crucial for personalized recommendation.

Second, this paradigm encounters a compositionality failure.
Because clients cannot directly access item attribute features under the dual-sided constraint, cold-item recommendation must inevitably rely on the server-side attribute encoder.
However, existing methods typically restrict this encoder to a single global mapping function, forcing all heterogeneous attribute signals to be embedded into one shared latent space.
This imposes a uniform interpretation of attributes across all items, even though different items may depend on different semantic facets.

The toy example again highlights this issue.
From a drug-oriented perspective, item $A$ may be more similar to item $B$ than to item $C$, since $A$ and $B$ both involve drug-related categories.
In contrast, from an ML/graph-oriented perspective, item $A$ may be more similar to item $C$, since $A$ and $C$ share methodological categories such as \emph{ML} and \emph{Graph}.
Formally, this implies that it is possible to have
$\mathrm{Sim}(A,B) > \mathrm{Sim}(A,C)$ under one semantic view,
while
$\mathrm{Sim}(A,B) < \mathrm{Sim}(A,C)$ under another.
A single encoder cannot faithfully preserve these conflicting neighborhood structures, because it is forced to collapse all semantic criteria into one representation space with one uniform notion of similarity.
As a consequence, heterogeneous semantics become entangled, item-to-item relations are distorted, and the quality of cold-item recommendation degrades.

Importantly, this issue cannot be resolved merely by introducing multiple view-specific encoders and na\"ively averaging their outputs.
Such averaging would again produce a uniform mixture of all semantic aspects, ignoring the fact that different items should emphasize different views.
In the toy example, item $B$ should place more weight on the drug/chemistry view, whereas item $C$ should place more weight on the ML/graph/recommendation view.
Therefore, what is needed is not only view disentanglement, but also an item-adaptive composition mechanism that can selectively emphasize the most relevant semantic views for each item.

\begin{figure}[t]
    \centering
    \includegraphics[width=1.0\linewidth]{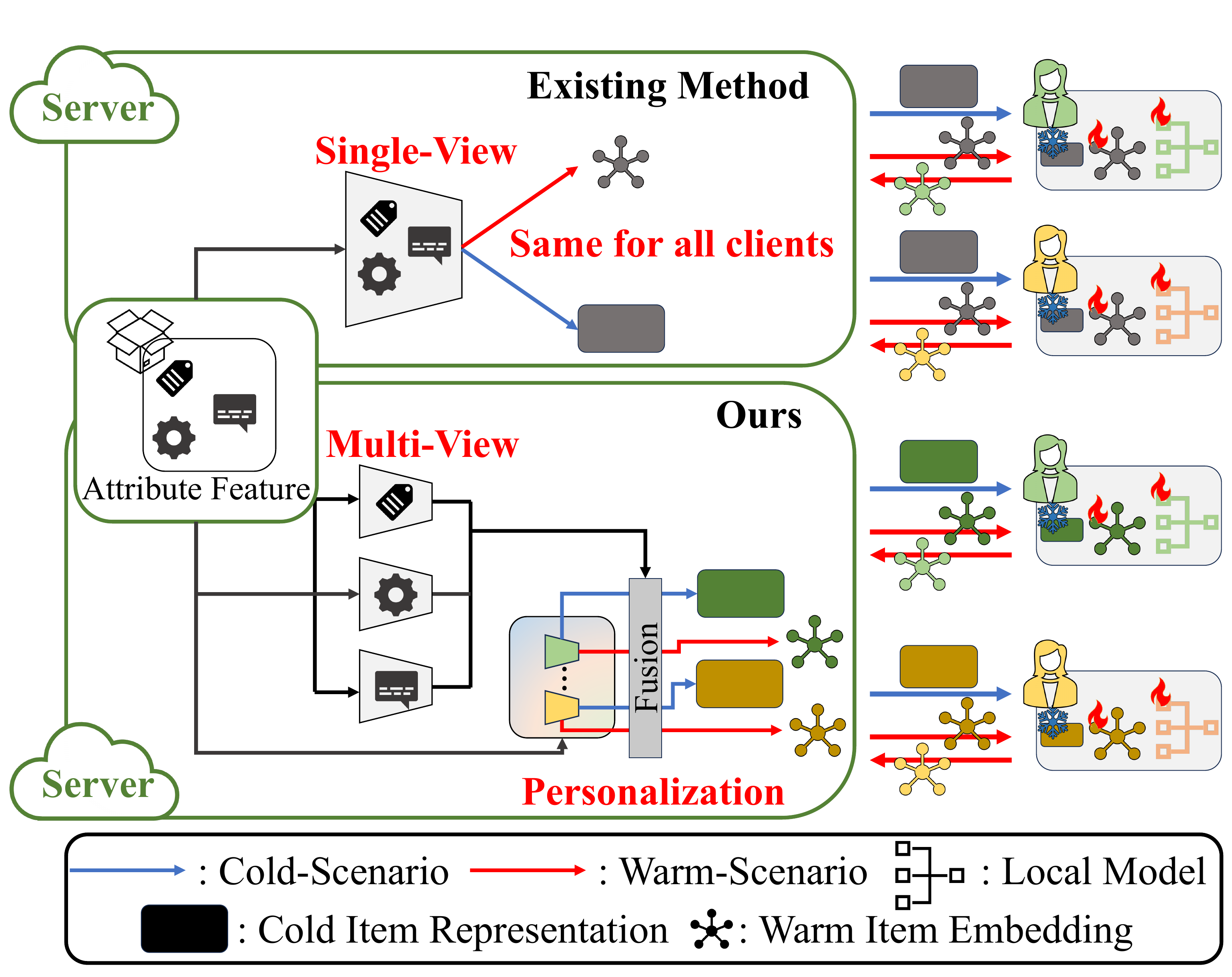}
    \caption{Comparison between existing methods and PMFRec. Prior federated cold-start approaches rely on a global single encoder, resulting in identical representations. In contrast, PMFRec introduces a personalized and multi-view encoding to generate user-specific representations for same cold items.}
    \label{fig:intro_concept}
\end{figure}

Third, existing dual-sided approaches are training- and communication-inefficient because they keep collaborative knowledge and attribute knowledge as separate representations and couple them only through an explicit alignment regularizer. As a result, clients must repeatedly receive both the warm-item embedding matrix and the attribute-feature-based item representation matrix during training. In large-scale recommendation settings with massive item pools, this dual-representation design imposes substantial communication and device-side memory overhead. Moreover, because the local objective must simultaneously optimize recommendation and cross-representation alignment, optimization becomes unnecessarily complicated. Therefore, a more desirable solution is to integrate collaborative and attribute knowledge into a single exchanged item representation and train it without requiring an explicit client-side regularizer.

To address these challenges, we propose \underline{\textbf{P}}ersonalized and \underline{\textbf{M}}ulti-view Representation for \underline{\textbf{F}}ederated Cold-Start \underline{\textbf{Rec}}ommendation (\textbf{PMFRec}). (Figure~\ref{fig:intro_concept} compares PMFRec with existing methods.)
First, to tackle the lack of personalization, we introduce an additional personalized encoder on the server under the dual-sided constraint.  As a result, each user can generate distinct cold-item representations that reflect their own preferences.
Second, to mitigate compositionality failure, PMFRec models heterogeneous attribute semantics with view-specific projectors and an item-wise gating mechanism, and further encourages view diversity via an orthogonality (disentanglement) objective.

To address these challenges, we propose \underline{\textbf{P}}ersonalized and \underline{\textbf{M}}ulti-view Representation for \underline{\textbf{F}}ederated Cold-Start \underline{\textbf{Rec}}ommendation (\textbf{PMFRec}) (Figure~\ref{fig:intro_concept}).
First, to tackle the lack of personalization, we introduce a personalized representation generator on the server under the dual-sided constraint, enabling user-specific cold-item representations.
Second, to mitigate compositionality failure, PMFRec models heterogeneous attribute semantics through view-specific projectors and an item-adaptive gating mechanism, while encouraging view diversity via an orthogonality objective.
Third, to avoid the inefficiency of dual-representation alignment, PMFRec fuses collaborative and attribute knowledge into a single exchanged item representation, eliminating the need for an explicit client-side regularizer.

Our contributions are summarized as follows:
\begin{itemize}
    \item We are the first to systematically introduce \emph{personalization} into federated cold-start recommendation under dual-sided constraint.
    \item We address the compositionality failure arising in global knowledge formation through a \emph{multi-view encoder}, an \emph{item-adaptive gating mechanism}, and an \emph{orthogonality objective}.
    \item We introduce a regularizer-free training scheme that integrates collaborative and attribute knowledge into a single exchanged item representation, thereby reducing communication overhead and avoiding explicit client-side alignment regularization.
    \item We validate the effectiveness of our method through extensive experiments on real-world datasets.
\end{itemize}

\section{Related Works}

\subsection{Federated Recommendation}
Federated recommendation (FedRec) enables collaborative model training without centralizing clients' raw interaction data, exchanging only model parameters or updates. Early work applies federated learning to recommendation \citep{fcf, fedmf}, and subsequent studies extend FedRec to neural recommenders \citep{fedncf}. Personalization is then explored by \citep{pfedrec}, which improves user-specific modeling under federated constraints. Other lines of research incorporate additional structures such as graphs \citep{gpfedrec, fedpergnn} or dual knowledge/representation learning \citep{feddae, fedrap}, or mitigate information loss from direct aggregation through alternative knowledge sharing mechanisms \citep{fedcia, fediar}. Despite these advances, most FedRec methods assume a fixed item pool and do not explicitly address new items.

\subsection{Cold-Start Recommendation}
\noindent \textbf{Centralized Cold-Start Recommendation.}
Cold-start recommendation aims to recommend new items with insufficient or no interaction history \citep{cold_cf_1, cold_cf_2, heater, hybrid_1, hybrid_2, gar, gorec}.
Existing approaches typically fall into three categories: CF methods leveraging historical interactions \citep{cold_cf_1, cold_cf_2}, content-based methods using item-side attributes \citep{content_1}, and hybrid models that integrate both interaction and content signals \citep{heater, hybrid_1, hybrid_2, gar, gorec}.
However, these centralized approaches assume that user interactions are collected at a server, raising privacy concerns and motivating FL solutions.

\noindent \textbf{Federated Cold-Start Recommendation.}
Most FedRec studies assume a fixed item pool, whereas real services continuously introduce new items, motivating federated cold-start recommendation.
\citep{first_fedcold} first points out cold-start issues in federated settings, but allows cold items to have interactions and does not leverage item attributes.
\citep{ifedrec} instead considers a more realistic setting with \emph{zero-interaction} cold items that must be recommended purely from attribute features under the \emph{dual-sided constraint}, while \citep{fedcolduser} studies cold users who do not join warm training.
Although \citep{fcrec} considers continual arrival of both new users and new items, it focuses on continual learning without attribute-based modeling for cold-start recommendation.
However, personalization for interaction-free cold items under the dual-sided constraint remains underexplored.
We fill this gap by enabling personalized representations for zero-interaction cold items in federated recommendation.

\section{Preliminaries}

\subsection{Problem Formulation}
\label{subsec:problem_formulation}
Let $\mathcal{U}$ denote the set of users, and partition the item set $\mathcal{I}$ into warm items $\mathcal{I}^{W}$ and cold items $\mathcal{I}^{C}$.
Each item $i$ is associated with a $d_f$-dimensional attribute feature vector $X_i$.
We collect the server-side attribute features of warm and cold items into matrices
$X^{W}\in\mathbb{R}^{|\mathcal I^{W}|\times d_f}$ and $X^{C}\in\mathbb{R}^{|\mathcal I^{C}|\times d_f}$, respectively.

\noindent \textbf{Client-side model.}
Each user $u\in\mathcal{U}$ is treated as a client and maintains private parameters
$\mathcal W_u=\{e_u,\Theta_u\}$, including a user embedding $e_u\in\mathbb{R}^{d}$ and parameters $\Theta_u$ of a personalized scoring function $\mathcal{F}_u(\cdot;\Theta_u)$.
User $u$ also maintains a private warm-item embedding matrix $Q_u\in\mathbb{R}^{|\mathcal{I}^{W}|\times d}$, where $d$ is the embedding dimension.

\noindent \textbf{Server-side attribute-feature-based item representations.}
Unlike prior FedRec methods that store a global item embedding matrix at the server, our server does \emph{not} maintain global item embeddings.
Instead, at each training round, the server produces item representations from $X^{W}$ and shares them to support client training.
For cold-item inference, the server similarly generates representations for cold items from $X^{C}$.

\noindent \textbf{Federated training objective.}
At each round $t$, the server samples a subset of users $\mathcal{U}^{t}\subseteq\mathcal{U}$ to participate in training.
Given the learned client-side parameters and the server-generated attribute-feature-based representations, our goal is to predict personalized preferences over cold items with no interaction records that are described only by their attribute features available only at the server.


\subsection{Dual-Sided Constraints}
\label{subsec:dual_sided_constraints}

Federated cold-start recommendation is constrained by privacy requirements on both the client and the server sides \citep{ifedrec}.

\noindent \textbf{Client privacy.}
To preserve user privacy, the server cannot access clients' interaction histories nor private user parameters,
including user embeddings and personalized scoring functions. Following FedRec-style protocols, the server is restricted
from observing $\{e_u,\Theta_u\}$ and the local interaction logs of any client $u$.

\noindent \textbf{Server confidentiality.}
From the server's perspective, fine-grained item attribute features are valuable commercial assets. Thus, the server does
not share item attribute features with clients. Consequently, clients do not have access to $X^{W}$ or $X^{C}$, and they
should not be able to infer item attribute features from exchanged messages.

\begin{figure*}
    \centering
    \includegraphics[width=\linewidth]{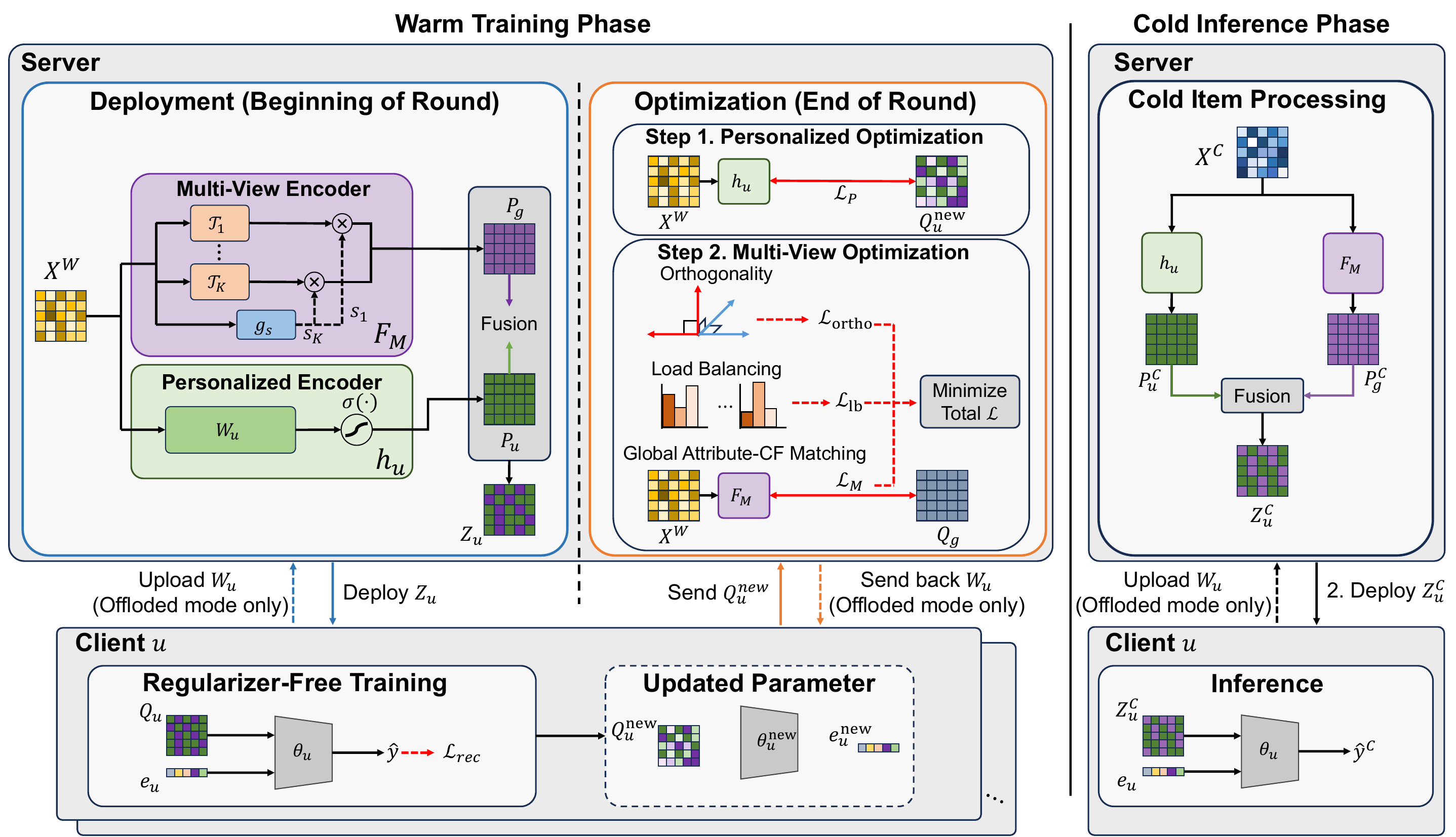}
    \caption{The overall framework of PMFRec.}
    \label{fig:method}
\end{figure*}


\section{Methodology}
\label{sec:methodology}
In this section, we present PMFRec, a framework that enables \emph{personalization} and \emph{multi-view mapping} for federated cold-start recommendation under \emph{dual-sided constraints}.
We first describe the regularizer-free warm training process, where collaborative knowledge is fused into server-provided attribute-feature-based representations without requiring client-side alignment regularization (Section~\ref{subsec:warm_client}).
We further show how the server aggregates client-updated item embeddings without storing all item embedding matrices from participating clients, enabling storage-efficient knowledge aggregation.
We then explain how the server uses the aggregated collaborative knowledge to learn (1) a personalized representation generator for each user and (2) a global multi-view encoder (Section~\ref{subsec:warm_server}).
Finally, we present the cold inference phase, where the server generates personalized and multi-view cold-item representations and clients perform inference (Section~\ref{subsec:cold_phase}).


\subsection{Warm Training Phase: Client-side}
\label{subsec:warm_client}
On the client side, each client learns its private parameters and warm-item embeddings from local interaction data. Our training design allows collaborative signals to be absorbed into attribute-feature-based initial representations through standard recommendation training, without introducing an explicit client-side regularizer. This also improves communication efficiency by requiring the exchange of only a single item representation matrix per round.

\subsubsection{\textbf{Client-side Regularizer-free Training}}
\label{subsubsec:client_objective}

At the beginning of each communication round, the server sends each participating client $u$ a warm item representation matrix $Z_u \in \mathbb{R}^{|I^{W}|\times d}$.
The client then initializes its warm-item embedding matrix $Q_u$ with $Z_u$.
Importantly, in our framework, $Z_u$ may differ across clients, and it is computed from warm-item
attribute features at the server (the construction of $Z_u$ is described in
Section~\ref{subsubsec:deploy_warm}).

\noindent \textbf{Local objective.}
Client $u$ optimizes a standard implicit-feedback objective (Binary Cross Entropy) over warm items
\begin{equation}
\label{eq:local_loss}
\mathcal L_u
=
-\sum_{i\in \mathcal D_{u}^{(W,+)}} y_{ui}\log \hat y_{ui}
-\sum_{i\in \mathcal D_{u}^{(W,-)}} (1-y_{ui})\log(1-\hat y_{ui}),
\end{equation}
where $\mathcal D_{u}^{(W,+)}\subseteq \mathcal I^{W}$ is the set of warm items interacted with by user $u$,
and $\mathcal D_{u}^{(W,-)}$ is a negative set sampled from $\mathcal I^{W}\setminus \mathcal D_{u}^{(W,+)}$.
The predicted score $\hat y_{ui}$ on item $i$ is computed by the personalized scoring function
\begin{equation}
\label{eq:score_fn}
\hat y_{ui} = \mathcal{F}_u(Q_{ui}, e_u;\Theta_u)
\end{equation}
where we implement $\mathcal{F}_u(\cdot;\Theta_u)$ as a two-layer MLP that takes the concatenation of $Q_{ui}$ and $e_u$ as input.
Under dual-sided constraints, the client does not observe $X^W$ or the server's attribute-feature-to-embedding mapping.
Conceptually, $Z_u$ is generated from attribute features via an unknown server function $\mathcal{R}(\cdot)$ (i.e.,
$Z_u = \mathcal{R}(X^W)$), but the client only receives $Z_u$ as an initialization.
Unlike prior methods that transmit two separate representations, our framework exchanges only a single type of item embedding at each communication round. This design is motivated by communication efficiency: sending two different item embeddings in every round incurs substantial overhead. In contrast, we show that exchanging only one embedding type is sufficient to achieve effective learning, while significantly reducing communication cost.

\noindent \textbf{Local Optimization.}
The client then performs local optimization by gradient descent
\begin{equation}
\label{eq:client_updates}
Q_u \leftarrow Q_u - \eta_Q \nabla_{Q_u}\mathcal L_u,
\quad
\mathcal W_u\leftarrow \mathcal W_u-\eta_{\mathcal W} \nabla_{\mathcal{W}_u} \mathcal L_u.
\end{equation}
Since local recommendation training is performed on top of attribute-feature-based initialization, collaborative knowledge is progressively absorbed into the initialized representation through optimization dynamics. In this way, PMFRec achieves collaborative-attribute fusion without requiring an alignment regularizer on the client side.

\subsection{Warm Training Phase: Server-side}
\label{subsec:warm_server}
On the server side, PMFRec uses client-updated warm-item embeddings to learn both user-specific and global attribute-based knowledge. Specifically, the server first performs storage-efficient aggregation to obtain global collaborative knowledge, while also constructing a personalized representation generator for each client from its uploaded warm-item embeddings. The aggregated collaborative knowledge is then used to train the global multi-view encoder. Together with the personalized representation generator constructed for each client, the global multi-view encoder produces client-specific item representations for subsequent training and cold-item inference.

\subsubsection{\textbf{Incremental Global Aggregation}}

After receiving the updated warm-item embedding matrix $Q_u$ from client $u$, the server incrementally aggregates the uploaded embeddings into a global warm-item embedding matrix. Specifically, at the beginning of each communication round $t$, the server initializes $Q_g$ to zero and updates it as each participating client's upload arrives:
\begin{equation}
    \label{eq:cum_agg}
    Q_g \leftarrow Q_g + \frac{1}{|\mathcal{U}^t|} Q_u .
\end{equation}
After all participating clients in $\mathcal{U}^t$ have been processed, this yields
\begin{equation}
    Q_g =
    \frac{1}{|\mathcal{U}^t|}
    \sum_{u\in\mathcal{U}^t} Q_u .
\end{equation}

This streaming aggregation avoids retaining all uploaded client embeddings in server memory. This is important in large-scale federated recommendation systems because clients may return their updates at different times due to heterogeneous local computation speeds, device conditions, and network environments. Therefore, instead of storing all $Q_u$ matrices until every client update arrives, the server sequentially accumulates each received $Q_u$ into $Q_g$.
The resulting $Q_g$ captures the shared collaborative filtering knowledge of the participating clients in round $t$ and is then used as the global target for learning the multi-view encoder (Section~\ref{subsubsec:multiview_mapping}).

\subsubsection{\textbf{Personalized Representation Generator}}

In addition to the global aggregation in Eq.~\eqref{eq:cum_agg}, the server uses each uploaded $Q_u$ to construct a user-specific personalized representation generator $h_u(\cdot):\mathbb{R}^{d_f}\rightarrow\mathbb{R}^{d}$. Since cold items have no interaction records and are available only through server-side attribute features, cold-item recommendation requires an attribute-to-representation mapping. However, even for the same item attribute features, different users may emphasize different aspects according to their warm-interaction patterns. PMFRec therefore fits $h_u(\cdot)$ to map warm-item attributes $X^W$ to the client-updated embedding matrix $Q_u$, which contains user-specific collaborative filtering knowledge learned from user $u$'s warm interactions. Once learned, $h_u(\cdot)$ can be applied to cold-item attributes $X^C$, enabling zero-interaction cold items to be interpreted from user-specific perspectives.

We implement $h_u(\cdot)$ as a lightweight linear mapping followed by a non-linear activation:
\begin{equation}
    h_u(X) = \sigma(XW_u),
\end{equation}
where $X \in \{X^W, X^C\}$ denotes the warm- or cold-item attribute feature matrix, $\sigma(\cdot)$ is a non-linear activation function, and $W_u \in \mathbb{R}^{d_f \times d}$ is the user-specific mapping parameter.
A straightforward way to fit $h_u(\cdot)$ is to minimize the following objective:
\begin{equation}
    \label{eq:l_p_original}
    \mathcal{L}_P
    =
    \|h_u(X^W) - Q_u\|_F^2
    +
    \lambda \|W_u\|_F^2 .
\end{equation}
However, directly optimizing Eq.~\eqref{eq:l_p_original} for every participating client would require maintaining a separate optimizer and performing iterative server-side updates for each user, which may become inefficient in large-scale systems. Therefore, we reformulate the objective into a ridge-regression form by applying the inverse activation to $Q_u$:
\begin{equation}
    \label{eq:l_p}
    \mathcal{L}_P
    =
    \|X^W W_u - \sigma^{-1}(Q_u)\|_F^2
    +
    \lambda \|W_u\|_F^2 .
\end{equation}
Then, Eq.~\eqref{eq:l_p} has the following closed-form solution:
\begin{equation}
    \label{eq:w_closed_form}
    W_u
    =
    \left((X^W)^{\top}X^W+\lambda I\right)^{-1}
    (X^W)^{\top}\sigma^{-1}(Q_u).
\end{equation}
This closed-form formulation avoids iterative optimization for each personalized encoder. Moreover, the factor preceding $\sigma^{-1}(Q_u)$ in Eq.~\eqref{eq:w_closed_form} depends only on $X^W$ and can therefore be precomputed once at the server. As a result, when a client uploads $Q_u$, the server can efficiently obtain $W_u$ by applying the inverse activation to $Q_u$ and multiplying it by the precomputed factor.

\paragraph{Deployment Modes for Scalability.}
Maintaining a personalized representation generator for every user can introduce additional server-side storage overhead when the number of users is very large. To address this issue, PMFRec supports two deployment modes.

\noindent\textbf{Server-retained mode.}
In this mode, the server stores the personalized mapping parameter $W_u$ for each user after fitting Eq.~\eqref{eq:w_closed_form}. When user $u$ participates in a later communication round or requests cold-item recommendation, the server reuses the stored $W_u$ to generate personalized warm- or cold-item representations.

\noindent\textbf{Storage-efficient offloaded mode.}
When server-side storage is limited, the server does not persistently store $W_u$. After using the uploaded $Q_u$ for incremental aggregation and personalized representation generator construction, the server sends the resulting $W_u$ back to client $u$ and discards both $Q_u$ and $W_u$ from server memory. When the same client participates in a future training round or requests cold-item inference, the client sends its stored $W_u$ to the server, which uses it temporarily to generate $h_u(X^W)$ or $h_u(X^C)$.

These two deployment modes provide a trade-off between server-side storage and communication cost.

\paragraph{Privacy and Dual-sided Constraint.}
The server does not observe the user's private parameters, such as the user embedding $e_u$ and the personalized scoring function $\mathcal F_u(\cdot;\Theta_u)$. Therefore, possessing $h_u(\cdot)$ alone does not allow the server to reconstruct the user's exact preference function. Meanwhile, in the storage-efficient offloaded mode, the client receives $W_u$, which is estimated from the server-side attribute matrix $X^W$. Thus, the offloaded parameter should not provide a direct way to recover the server-side item attributes.

Our inverse pre-activation formulation provides an additional safeguard in this respect. Specifically, PMFRec estimates $W_u$ through the ridge-regularized objective in Eq.~\eqref{eq:l_p} with $\lambda>0$.
Because $W_u$ is obtained through this regularized pre-activation mapping, it is generally not an exact interpolation operator satisfying $X^W W_u=\sigma^{-1}(Q_u)$. Hence, the simple algebraic inversion that could arise from an exact unregularized mapping is not directly applicable. This provides a practical safeguard against direct linear inversion while enabling the server to offload $W_u$ for better scalability.

\subsubsection{\textbf{Multi-view Encoder}}
\label{subsubsec:multiview_mapping}

Personalized knowledge alone can be unstable or insufficient, so we also learn a \emph{global multi-view encoder} that provides a global perspective.
Existing federated approach relies on a single attribute-feature-to-embedding mapping, which can suffer from compositionality failures when heterogeneous semantics are forced into one space. We therefore decompose attribute knowledge into $K$ views.

Let $\mathcal T_k(\cdot)$ be the $k$-th view encoder that maps an item feature vector to the embedding space,
$\mathcal T_k:\mathbb{R}^{d_f}\rightarrow\mathbb{R}^{d}$ (implemented as a one-layer MLP).
We define the global multi-view representation generator $F_M(\cdot)$ as
\begin{equation}
\label{eq:FM_def}
F_M(X_i^W)=\sum_{k=1}^{K} s_{ik}\, \mathcal T_k(X_i^W),
\end{equation}
where $s_{ik}$ is the view weight for item $i$, computed by
\begin{equation}
\label{eq:view_weights}
s_{ik}=\frac{\exp\!\left(g_s(X_i^W)_k\right)}{\sum_{k'=1}^{K}\exp\!\left(g_s(X_i^W)_{k'}\right)}.
\end{equation}
Here, $g_s(\cdot)$ denotes the gating network for routing
\begin{equation}
\label{eq:gate_linear}
g_s(X_i^W)= X_i^W W_s + b_s \in\mathbb{R}^{K},
\end{equation}
where $W_s\in\mathbb{R}^{d_f\times K}$ and $b_s\in\mathbb{R}^{K}$.
Each $\mathcal T_k(\cdot)$ serves as a view-specific projector that interprets the same attribute feature from a different
semantic perspective and maps it into a embedding space.
Consequently, a single attribute feature can yield multiple view-dependent representations, which are then combined by the gating weights $\{s_{ik}\}_{k=1}^K$ to form the multi-view representation.

\noindent \textbf{Global Attribute-CF Matching.}
Since client uploads arrive sequentially, the server incrementally accumulates the warm-item embeddings received from the participating users in round $t$. After all users in $\mathcal U^t$ have uploaded their updated item embeddings, the resulting matrix forms the global warm-item embedding matrix:
\begin{align}
    \label{eq:aggregate}
    Q_g = \frac{1}{|\mathcal U^t|}\sum_{u\in\mathcal U^t} Q_u.
\end{align}
This aggregated $Q_g$ can be viewed as capturing the shared collaborative filtering knowledge across the participating users. The server then aligns the multi-view attribute mapping to this global CF knowledge by minimizing
\begin{equation}
\label{eq:LM}
\mathcal L_M = \left\|F_M(X^W)-Q_g\right\|_F^2.
\end{equation}

\noindent \textbf{Orthogonality.}
Simply using multiple views does not guarantee that the view subspaces capture non-redundant information.
To reduce redundancy, we define view-level knowledge vectors
\begin{equation}
\label{eq:Jk}
J_k=\frac{1}{|\mathcal I^W|}\sum_{i\in\mathcal I^W} \mathcal T_k(X_i^W),
\end{equation}
and construct $A=[\bar J_1,\ldots,\bar J_K]\in\mathbb{R}^{d\times K}$ with $\bar J_k=J_k/\|J_k\|_F$.
We define the orthogonality loss as
\begin{equation}
\label{eq:L_ortho}
\mathcal L_{\mathrm{ortho}}
=
\frac{1}{K(K-1)}\left\|A^\top A - I\right\|_F^2
=
\frac{1}{K(K-1)}
\sum_{\substack{1\le n,m\le K\\ n\ne m}}
\left(\frac{J_n^\top J_m}{\|J_n\|_F\,\|J_m\|_F}\right)^2.
\end{equation}
Minimizing $\mathcal L_{\mathrm{ortho}}$ encourages the view to be orthogonal on average, thereby reducing redundancy among the knowledge captured by different views.
This makes the multi-view decomposition meaningful, as each view is pushed to encode distinct information.

\noindent \textbf{Load Balancing.}
To prevent view collapse (over-selecting a subset of views), we further apply load balancing. Let
\begin{equation}
\label{eq:L_lb}
\mathcal L_{\mathrm{lb}}=-\sum_{k=1}^{K} S_k \log(S_k+\xi), \quad S_k=\frac{1}{|\mathcal I^W|}\sum_{i\in\mathcal I^W} s_{ik},
\end{equation}
where $\xi>0$ is a small constant. Here, $S_k$ is the average routing weight assigned to the $k$-th view over all warm items. By optimizing $\mathcal L_{\mathrm{lb}}$, the gate encourages balanced view usage in aggregate.

\noindent \textbf{Final Server-side Objective.}
The server optimizes the multi-view encoder and gating network by minimizing
\begin{equation}
\label{eq:server_total_loss}
\mathcal L = \mathcal L_M  + \lambda_{\mathrm{ortho}}\mathcal L_{\mathrm{ortho}} - \lambda_{\mathrm{lb}}\mathcal L_{\mathrm{lb}}.
\end{equation}

\subsubsection{\textbf{Parameter Deployment in the Warm Training Phase}}
\label{subsubsec:deploy_warm}
We now describe how the server constructs the client-specific initialization $Z_u$ used in
Section~\ref{subsubsec:client_objective}.
At the start of round $t (>1)$, the server generates a personalized and multi-view warm-item representation:
\begin{equation}
\label{eq:Zu}
Z_u = \frac{1}{2}\left(h_u(X^W) + F_M(X^W)\right).
\end{equation}

The server sends $Z_u$ to client $u$ and, for memory efficiency, discards temporary variables such as
$\{Z_u,P_u\}$ after transmission. This deployment ensures that the client learns CF knowledge from a representation that already fuses attribute-based
(global) knowledge and user-specific (personalized) knowledge, without requiring the client to optimize an explicit regularization term.


\subsection{Cold Inference Phase}
\label{subsec:cold_phase}

In the cold phase, we generate cold-item representations using the mappings learned in the warm training phase and then perform inference for each user.
Unlike prior work that forces all users to share the same cold-item view, our approach uses a user-specific encoder, allowing different users to interpret cold items through personalized perspectives.

Given a client $u$ who requests recommendations, the server constructs a user-specific and multi-view cold-item representation by
\begin{equation}
\label{eq:Zc}
Z_u^{C}=\frac{1}{2}\left(h_u(X^{C}) + F_M(X^{C})\right).
\end{equation}
The server delivers $Z_u^{C}$ to client $u$, enabling user-specific cold-start inference.
For a cold item $i\in\mathcal I^{C}$, the predicted preference is computed as $\hat{y}^C_{ui} = \mathcal{F}_u(Z_{ui}^C,e_u |\Theta_u)$.

\subsection{Additional Privacy Preservation}
\label{sec:ldp}
Most FedRec methods perform local training with negative sampling \cite{pfedrec,gpfedrec,cofedrec,fedrap,fediar,ifedrec}, so a single round of updates does not deterministically reveal a client’s positive items. However, across rounds, consistent per-item update patterns may still leak likely positives as negatives change over time. Accordingly, when stricter privacy is required, we can incorporate $(\varepsilon,\delta)$-LDP (Local Differential Privacy) on the client-released item-gradient matrix to mitigate both \emph{index leakage} and \emph{update-pattern leakage} \cite{ldp}. Notably, many existing FedRec approaches apply noise within a fixed absolute range without considering the magnitude of gradients or individual item embeddings. This oversight potentially allows an adversary to identify trained items by monitoring update variations in practical scenarios. To mitigate such risks and support enhanced privacy-preservation scenarios, we integrate the following mechanism.\\

\noindent \textbf{Local Differential Privacy.}
A randomized mechanism $\mathcal{M}$ satisfies $(\varepsilon,\delta)$-LDP if for any two inputs $\psi,\psi'$ and any measurable set $\mathcal{H}$,
$\Pr[\mathcal{M}(\psi)\in\mathcal{H}] \le e^\varepsilon \Pr[\mathcal{M}(\psi')\in\mathcal{H}] + \delta,$
where $\varepsilon$ is the privacy budget and $\delta$ is the allowable failure probability.

To enforce $(\varepsilon,\delta)$-LDP on client updates, each local step privatizes the item-embedding gradient
$G=\nabla_{Q_u}\mathcal{L}_u \in \mathbb{R}^{|\mathcal I^W|\times d}$ by clipping and adding Gaussian noise:
\begin{align}
\label{eq:ldp}
\tilde G = \mathrm{clip}_C(G) + Z,
\end{align}
where $\mathrm{clip}_C(G)\triangleq \min\!(1,\frac{C}{\|G\|_F})G$ and
$Z\in\mathbb{R}^{|\mathcal I^W|\times d}$ is sampled such that
$\mathrm{vec}(Z)\sim\mathcal{N}\!(0,\sigma^2 C^2 I_{|\mathcal I^W|d})$.
Here, $\mathrm{vec}(\cdot)$ stacks matrix columns into a vector and $I_m$ denotes $m\times m$ identity matrix.
The client then updates the item embeddings as
\begin{align}
\label{eq:ldp_local_update}
Q_u^{\mathrm{new}} \leftarrow Q_u - \eta\,\tilde G .
\end{align}
Since the server only observes the communicated change $\Delta Q_u=-\eta\tilde G$, the per-round embedding update is privatized.
We use the standard Gaussian calibration to set $\sigma$ for the given $(\varepsilon,\delta)$. A proof that this mechanism satisfies $(\varepsilon,\delta)$-LDP is provided in Appendix~\ref{appendix:ldp}.


\section{Experiments}
In this section, we evaluate PMFRec through the following research questions (RQs):
\textbf{RQ1}. whether PMFRec outperforms prior methods;
\textbf{RQ2}. how each component contributes to performance (ablation);
\textbf{RQ3}. whether personalization mitigates preference dilution and improves user-level outcomes;
\textbf{RQ4}. whether the multi-view design learns diverse and well-utilized views; 
\textbf{RQ5}. whether our personalized and multi-view representations can be integrated into existing baselines and support warm-item training; and
\textbf{RQ6}. how PMFRec converges and how sensitive it is to hyperparameters;
\textbf{RQ7}. how PMFRec behaves under LDP.

\begin{table}[t]
    \small
    \centering
    \begingroup
    \renewcommand{\arraystretch}{0.92}
    \setlength{\tabcolsep}{4pt}
    \setlength{\aboverulesep}{0.2ex}
    \setlength{\belowrulesep}{0.2ex}
    \setlength{\cmidrulesep}{0.2ex}
    \begin{tabular}{llcccc}
        \toprule
        Dataset & Split & \#Users & \#Items & \#Interactions & Sparsity \\
        \midrule
        \multirow{3}{*}{CiteULike}
          & Train & 5,551 & 13,584 & 164,210 & 99.78\% \\
          & Val   & 5,551 & 1,018  & 13,037  & -- \\
          & Test  & 5,551 & 2,378  & 27,739  & -- \\
        \midrule
        \multirow{3}{*}{XING-5000}
          & Train & 5,000 & 11,290 & 120,944 & 99.79\% \\
          & Val   & 5,000 & 1,869  & 21,713  & -- \\
          & Test  & 5,000 & 5,648  & 58,609  & -- \\
        \midrule
        \multirow{3}{*}{XING-10000}
          & Train & 10,000 & 12,144 & 241,595 & 99.80\% \\
          & Val   & 10,000 & 2,018  & 43,904  & -- \\
          & Test  & 10,000 & 6,068  & 117,797 & -- \\
        \midrule
        \multirow{3}{*}{XING-20000}
          & Train & 20,000 & 12,306 & 485,199 & 99.80\% \\
          & Val   & 20,000 & 2,051  & 88,330  & -- \\
          & Test  & 20,000 & 6,154  & 236,361 & -- \\
        \bottomrule
    \end{tabular}
    \endgroup
    \caption{Statistics of the four cold-start recommendation datasets. Sparsity is reported for the training split.}
    \label{tab:dataset}
\end{table}

\subsection{Experimental Setup}

\subsubsection{Datasets.}
We evaluate PMFRec on two widely used real-world cold-start recommendation datasets, \textbf{CiteULike} \cite{citeulike} and \textbf{XING} \cite{xing}, both of which provide rich item-side attribute information; detailed statistics are summarized in Table~\ref{tab:dataset}. \textbf{CiteULike} is an online article recommendation dataset. Each article is associated with a title and abstract, which are used as item-side information. Following \cite{heater, ifedrec}, we extract 8,000-dimensional TF--IDF features and apply SVD to obtain 300-dimensional attribute features. For evaluation, items are split into warm and cold sets at an 80/20 ratio, and the cold items are further divided into validation and test sets with a 30/70 ratio. \textbf{XING}, from the ACM RecSys 2017 Challenge, with each item represented by 2,738-dimensional attribute features. Following \cite{ifedrec}, we construct three user-sampled subsets, \textbf{XING-5000}, \textbf{XING-10000}, and \textbf{XING-20000} and adopt the same warm/cold split protocol as in \cite{heater, ifedrec}. Specifically, items are divided into warm, validation, and test sets with a 6:1:3 ratio.

\subsubsection{Baselines.}
We summarize the baselines as follows:
\begin{itemize}
    \item \textbf{FedMVMF} \cite{fedmvmf}: A federated MF model that combines interactions with item attribute features. Clients keep user-specific factors locally, while the server updates shared components.

    \item \textbf{FedWDR} \cite{wdr}: A federated Wide \& Deep Recommendation model. Clients train locally, and the server synchronizes shared wide/deep parameters.

    \item \textbf{FedNCF\_CS} \cite{fedncf}: A cold-start variant of FedNCF where item embeddings are replaced by a one-layer MLP that maps attribute features to item representations.

    \item \textbf{PFedRec\_CS} \cite{pfedrec}: A cold-start variant of PFedRec where item embeddings are replaced by a one-layer attribute-feature-to-representation MLP.

    \item \textbf{FedVBPR} \cite{vbpr}: A federated adaptation of VBPR, which augments CF with content features.

    \item \textbf{FedDCN} \cite{dcn}: A federated Deep \& Cross Network; clients train user-specific parts locally, while the server synchronizes shared embedding and cross-network parameters.

    \item \textbf{IFedNCF} \& \textbf{IPFedRec} \cite{ifedrec}: Baselines under the IFedRec framework with dual-sided constraints. The server encodes item attribute features into representations and applies an alignment mechanism for cold-start; instantiations are based on FedNCF and PFedRec.
\end{itemize}

\subsubsection{Evaluation Metrics.}
Following \cite{heater, ifedrec}, we rank cold items for each user by predicted scores and report top-$K$ metrics Recall@$K$, Precision@$K$, NDCG@$K$.
For each user $u$, let $GT(u)$ be the ground-truth positives in the test set (cold items) and let $\mathrm{TopK}(u)$ be the top-$K$ ranked items.
We compute
\begin{align}
\mathrm{Recall@K} &= \frac{1}{|U|}\sum_{u\in U}
\frac{\left| \mathrm{TopK}(u) \cap GT(u) \right|}{|GT(u)|}, \\
\mathrm{Precision@K} &= \frac{1}{|U|}\sum_{u\in U}
\frac{\left| \mathrm{TopK}(u) \cap GT(u) \right|}{K}.
\end{align}
We also report NDCG@$K$:
\begin{align}
\mathrm{DCG@K}(u) &= \sum_{j=1}^{K} \frac{\mathrm{rel}_{u,j}}{\log_2(j+1)}, \\
\mathrm{IDCG@K}(u) &= \sum_{j=1}^{\min(K, |GT(u)|)} \frac{1}{\log_2(j+1)}, \\
\mathrm{NDCG@K} &= \frac{1}{|U|}\sum_{u\in U} \frac{\mathrm{DCG@K}(u)}{\mathrm{IDCG@K}(u)},
\end{align}
where $\mathrm{rel}_{u,j}=1$ if the item at rank $j$ is in $GT(u)$, and $0$ otherwise.

\subsubsection{Implementation Details}
All training and inference are conducted in PyTorch~\cite{pytorch} with CUDA on an RTX3090 GPU and an AMD EPYC 7313 CPU.
For our method and all baselines, we set the latent dimension to 64 and fix the batch size to 256.
We adopt negative sampling with a ratio of 5 negatives per positive item for all experiments.
The user sampling ratio per communication round is 1.0.
The number of global communication rounds is set to 300 for CiteULike and 400 for the XING datasets, while other baselines are run for 1000 rounds.
We set the number of local epochs to 1, and fix the number of server-side optimization epochs to 1 following \cite{ifedrec}.

\noindent \textbf{PMFRec.}
We use the SGD optimizer and tune the learning rate in $\{0.05, 0.1, 0.5, 1.0\}$ and server learning rate in $\{5\times10^{-4}, 10^{-3}, 5\!\times\!10^{-3}, 10^{-2}\}$.
We tune $\lambda_{\mathrm{lb}} \in \{10^{-4}, 10^{-3}, 10^{-2}, 10^{-1}\}$ and $\lambda_{\mathrm{ortho}} \in \{10^{-4}, 10^{-3}, 10^{-2}, 10^{-1}\}$.
The number of views is chosen from $\{4, 8, 16\}$ for all datasets.

\noindent \textbf{Baseline Configurations.}
For \textbf{FedMVMF}, we tune the learning rate in $\{0.1, 0.5, 1.0\}$ and the regularization coefficient in $\{0.1, 0.5, 1.0\}$.
For \textbf{FedWDR}, we treat the combination layer as private parameters while sharing the remaining parameters across clients; we tune the learning rate in $\{0.01, 0.05, 0.1, 0.5, 1.0\}$.
For the deep component, we tune the number of layers in $\{1, 2, 5\}$.
For \textbf{FedNCF\_CS}, we use a 2-layer MLP as the score function and tune the learning rate in $\{0.01, 0.05, 0.1, 0.5, 1.0\}$.
The client-side personalized feature mapping function is implemented as a 1-layer MLP.
For \textbf{PFedRec\_CS}, we use a 1-layer MLP as the score function and tune the learning rate in $\{0.01, 0.05, 0.1, 0.5, 1.0\}$;
the client-side personalized feature mapping function is also a 1-layer MLP.
For \textbf{FedVBPR}, we tune the learning rate in $\{0.01, 0.05, 0.1, 0.5, 1.0\}$.
For \textbf{FedDCN}, we tune the learning rate in $\{10^{-4}, 5\!\times\!10^{-4}, 10^{-3}, 5\!\times\!10^{-3}, 10^{-2}, 5\!\times\!10^{-2}, 10^{-1}, 0.5, 1.0\}$,
and tune the number of cross layers in $\{1, 2, 5\}$ as well as the number of deep layers in $\{1, 2, 5\}$.
For \textbf{FedDCN}, we treat only the score function as \emph{private} parameters and share the remaining parameters.
For \textbf{IFedNCF}, we tune the client learning rate in $\{0.05, 0.1, 0.5, 1.0\}$ and the regularization coefficient in $\{10^{-4}, 10^{-3}, 10^{-2}, 0.1, 0.5, 1.0, 10\}$.
We use a 2-layer MLP as the score function.
We additionally tune the server learning rate $\mathrm{lr}_{\mathrm{server}}$ in $\{10^{-3}, 5\!\times\!10^{-3}, 10^{-2}\}$.
For \textbf{IPFedRec}, we tune the client learning rate in $\{0.05, 0.1, 0.5, 1.0\}$ and the regularization coefficient in $\{10^{-4}, 10^{-3}, 10^{-2}, 0.1, 0.5, 1.0, 10\}$.
We use a 1-layer MLP as the score function, and tune the server learning rate $\mathrm{lr}_{\mathrm{server}}$ in $\{5\times10^{-4}, 10^{-3}, 5\!\times\!10^{-3}, 10^{-2}\}$.

\noindent \textbf{Implementation with LDP}
We conduct experiments under $(\varepsilon,\delta)$-LDP. We vary the clipping threshold (sensitivity) $C$ over
$\{10^{-5}, 10^{-3},$ $ 10^{-2}, 10^{-1}, 1\}$.
For IFedNCF and IFedNCF \textit{w/} PM, we tune the client-side regularization weight $\lambda_{\mathrm{reg}}$ in
$\{10^{-4}, 10^{-3}, 10^{-2}, 10^{-1}, 1, 10, 100\}$.
For model-level comparisons, we evaluate privacy budgets $\varepsilon \in \{4, 8, 10\}$, and set
$\delta = 1/|\mathcal U|^{1.1}$ for all experiments.
For IFedNCF \textit{w/} PM and PMFRec share the same hyperparameter configuration for the remaining settings
(e.g., $\eta$, $\eta_{\text{server}}$, $K$, $\lambda_{\mathrm{lb}}$, and $\lambda_{\mathrm{ortho}}$).

\begin{table*}[t]
\caption{Performance comparison across four datasets: the best results are highlighted in bold, specifically used when our model's result is larger than the best-performing baseline where $Improv(\%)$ represents the relative performance improvement over the best-performing baseline, and $*$ denotes statistical significance ($p\leq0.05$) for the paired t-test against the best baseline on each metric.}
\label{tab:main_full}
\resizebox{\linewidth}{!}{
\begin{tabular}{l|l|ccc|ccc|ccc|ccc}
\toprule
\multirow{2}{*}{Methods}& \multirow{2}{*}{Metrics} & \multicolumn{3}{c|}{CiteULike} & \multicolumn{3}{c|}{XING-5000} & \multicolumn{3}{c|}{XING-10000} & \multicolumn{3}{c}{XING-20000} \\

& & Recall & Precision & NDCG & Recall & Precision & NDCG & Recall & Precision & NDCG & Recall & Precision & NDCG \\

\midrule

\multirow{3}{*}{FedMVMF}        & @5& 0.1378& 0.1078& 0.1721& 0.0083& 0.0193& 0.0181& 0.0031& 0.0076& 0.0066& 0.0067& 0.0157& 0.0144\\
                                & @10& 0.2228& 0.0927& 0.2154& 0.0144& 0.0168& 0.0169& 0.0054& 0.0066& 0.0064& 0.0114& 0.0136& 0.0139\\
                                & @20& 0.3383& 0.0751& 0.2670& 0.0337& 0.0191& 0.0307& 0.0221& 0.0108& 0.0286& 0.0239& 0.0140& 0.0213\\

\midrule

\multirow{3}{*}{FedWDR}& @5& 0.1585& 0.1314& 0.2049& 0.0092& 0.0229& 0.0227& 0.0123& 0.0281& 0.0276& 0.0140& 0.0350& 0.0318\\
                                & @10& 0.2507& 0.1080& 0.2478& 0.0173& 0.0215& 0.0230& 0.0213& 0.0248& 0.0263& 0.0236& 0.0290& 0.2960\\
                                & @20& 0.3728& 0.0853& 0.2978& 0.0385& 0.0235& 0.0350& 0.0417& 0.0245& 0.0366& 0.0436& 0.0226& 0.0390\\

\midrule

\multirow{3}{*}{FedNCF\_CS}& @5& 0.1614& 0.1349& 0.2126& 0.0551& 0.1185& 0.1158& 0.0374& 0.0873& 0.0846& 0.0309& 0.0742& 0.0703\\
                                & @10& 0.2451& 0.1085& 0.2467& 0.0830& 0.0913& 0.0975& 0.0643& 0.0751& 0.0792& 0.0509& 0.0608& 0.0620\\
                                & @20& 0.3588& 0.0840& 0.2920& 0.1312& 0.0732& 0.1127& 0.1120& 0.0650& 0.0960& 0.0907& 0.0538& 0.0785\\

\midrule

\multirow{3}{*}{PFedRec\_CS}& @5& 0.1478& 0.1218& 0.1931& 0.0118& 0.0292& 0.0248& 0.0129& 0.0297& 0.0081& 0.0105& 0.0251& 0.0222\\
                                & @10& 0.2435& 0.1028& 0.2391& 0.0193& 0.0236& 0.0229& 0.0224& 0.0260& 0.0089& 0.0191& 0.0229& 0.0226\\
                                & @20& 0.3654& 0.0819& 0.2911& 0.0230& 0.0310& 0.0495& 0.0471& 0.0271& 0.0104& 0.0401& 0.0237& 0.0349\\

\midrule

\multirow{3}{*}{FedVBPR}& @5& 0.1373& 0.1160& 0.1800& 0.0494& 0.1140& 0.1075& 0.0502& 0.1169& 0.1096& 0.0332& 0.0799& 0.0722\\
                                & @10& 0.2258& 0.0992& 0.2257& 0.0891& 0.1028& 0.1055& 0.0910& 0.1058& 0.1106& 0.0627& 0.0747& 0.0750
\\
                                & @20& 0.3340& 0.0770& 0.2706& 0.1570& 0.0908& 0.1345& 0.1572& 0.0913& 0.1357& 0.1089& 0.0641& 0.0920\\

\midrule

\multirow{3}{*}{FedDCN}& @5& 0.1070& 0.0860& 0.1361& 0.0595& 0.1352& 0.1329& 0.0616& 0.1396& 0.1379& 0.0593& 0.1364& 0.1338\\
                                & @10& 0.1728& 0.0728& 0.1690& 0.1080& 0.1222& 0.1332& 0.1085& 0.1232& 0.1344& 0.1067& 0.1217& 0.1321\\
                                & @20& 0.2595& 0.0578& 0.2062& 0.1826& 0.1029& 0.1566& 0.1804& 0.1025& 0.1576& 0.1776& 0.1015& 0.1555\\

\midrule

\multirow{3}{*}{IFedNCF}        & @5& \underline{0.1817}& \underline{0.1521}& 0.2380& \underline{0.0868}& \underline{0.1939}& \underline{0.1901}&                                            \underline{0.0841}& \underline{0.1896}& \underline{0.1883}& \underline{0.0830}& \underline{0.1879}& \underline{0.1855}\\

                                & @10& 0.2774& 0.1211& 0.2763& \underline{0.1429}& \underline{0.1616}& \underline{0.1726}& \underline{0.1438}& \underline{0.1624}& \underline{0.1775}& \underline{0.1462}& \underline{0.1638}& \underline{0.1742}\\
                                            
                                & @20& \underline{0.4053}& \underline{0.0929}& \underline{0.3249}& \underline{0.2406}& \underline{0.1344}& \underline{0.2012}& \underline{0.2345}& \underline{0.1316}& 
                                            \textbf{0.2070}& \underline{0.2353}& \underline{0.1327}& \underline{0.2022}\\

\midrule

\multirow{3}{*}{IPFedRec}       & @5& 0.1798& 0.1520& \underline{0.2396}& 0.0813& 0.1829& 0.1794& 0.0771& 0.1747& 0.1724& 0.0753& 0.1719& 0.1697\\
                                & @10& \underline{0.2791}& \underline{0.1217}& \underline{0.2792}& 0.1372& 0.1550& 0.1650& 0.1304& 0.1485& 0.1616& 0.1295& 0.1486& 0.1604\\
                                & @20& 0.3940& 0.0907& 0.3160& 0.2353& 0.1311& 0.1987& 0.2117& 0.1213& 0.1854& 0.2230& 0.1264& 0.1893\\

\midrule

\multirow{3}{*}{Ours}& @5& \textbf{0.1921}*& \textbf{0.1621}*& \textbf{0.2549}*& \textbf{0.0872}*& \textbf{0.1952}*& \textbf{0.1948}*& \textbf{0.0853}*& \textbf{0.1917}*&  \textbf{0.1943}*& \textbf{0.0867}*& \textbf{0.1960}*& \textbf{0.1964}*\\

                                & @10& \textbf{0.2929}*& \textbf{0.1309}*& \textbf{0.3001}*& \textbf{0.1530}*& \textbf{0.1718}*& \textbf{0.1850}*& \textbf{0.1507}*& \textbf{0.1691}*& \textbf{0.1839}*& \textbf{0.1560}*& \textbf{0.1739}*& \textbf{0.1873}*\\
                                            
                                & @20& \textbf{0.4187}*& \textbf{0.0979}*& \textbf{0.3441}*& \textbf{0.2545}*& \textbf{0.1411}*& \textbf{0.2112}*& \textbf{0.2372}*& \textbf{0.1337}*& \underline{0.2057}& \textbf{0.2438}*& \textbf{0.1372}*& \textbf{0.2092}*\\
\midrule
\multirow{3}{*}{\textit{Improv}}& @5& 5.72\%& 6.57\%& 6.39\%& 0.46\%& 0.67\%& 2.47\%& 2.50\%& 2.16\%&  4.41\%& 4.46\%& 4.31\%& 5.88\%\\

                                & @10& 4.94\%& 7.56\%& 7.49\%& 7.07\%& 6.31\%& 7.18\%& 4.52\%& 3.82\%& 3.32\%& 6.70\%& 6.17\%& 7.52\%\\
                                            
                                & @20& 3.31\%& 5.38\%& 5.91\%& 5.78\%& 4.99\%& 4.97\%& 1.28\%& 1.60\%& -0.63\%& 3.61\%& 3.39\%& 3.46\%\\

\bottomrule
\end{tabular}%
}
\end{table*}

\subsection{Main Results (RQ1)}
Table~\ref{tab:main_full} reports the top-$K$ ranking performance across four datasets. We compare PMFRec with representative FedRec baselines and two state-of-the-art methods specifically designed for the dual-sided constraint setting: IFedNCF and IPFedRec. PMFRec achieves the best performance on nearly all dataset-metric pairs, demonstrating statistically significant improvements over the strongest baseline in most cases ($p \le 0.05$). The relative gains over the best-performing baseline reach up to 7.56\%, showcasing stable effectiveness across datasets with diverse scales and sparsity levels.

Notably, the improvements are consistent across Recall, Precision, and NDCG. This indicates that PMFRec enhances both retrieval quality and ranking fidelity in the cold-item setting. More importantly, even though PMFRec operates under strict dual-sided constraints, it consistently outperforms baselines that can directly exploit item attribute features, such as FedNCF\_CS, PFedRec\_CS, FedVBPR, and FedDCN. This result suggests that direct access to attribute features alone is insufficient for effective cold-start recommendation unless the model can also preserve user-specific preference signals. By jointly modeling personalization and multi-view representations, PMFRec effectively alleviates the \textsc{representation averaging} issue inherent in global feature-side mappings and provides more discriminative cold-item representations. Furthermore, the consistent improvements observed on the larger XING datasets support the scalability and robustness of our proposed framework in challenging cold-item recommendation scenarios.

\begin{table*}[h]
\scriptsize
\caption{Ablation study of PMFRec on CiteULike and XING-5000.}
\centering
\label{tab:ablation_full}
\resizebox{1.0\linewidth}{!}{%
    \begin{tabular}{l|cccccc|cccccc}
        \toprule
        \multirow{2}{*}{Methods} & \multicolumn{6}{c|}{CiteULike} & \multicolumn{6}{c}{XING-5000}  \\
          & R@10 & R@20 & P@10 & P@20 & N@10 & N@20 & R@10 & R@20 & P@10 & P@20 & N@10 & N@20 \\
        \midrule 
        PMFRec & \textbf{0.2929} & \textbf{0.4187} & \textbf{0.1309} & \textbf{0.0979} & \textbf{0.3001} & \textbf{0.3441} & \textbf{0.1530} & \textbf{0.2545} & \textbf{0.1718} & \textbf{0.1411}  & \textbf{0.1850} & \textbf{0.2112}  \\
         \textit{w/o} $P_u$ & 0.2675 & 0.3900 & 0.1200 & 0.0916 & 0.2763 & 0.3234 & 0.1367 & 0.2304 & 0.1553 & 0.1294 & 0.1662 & 0.1952  \\
         \textit{w/ only} $P_u$ & 0.2656 & 0.3988 & 0.1179 & 0.0922 & 0.2739 & 0.3244 & 0.1181 & 0.1927 & 0.1334 & 0.1093 & 0.1395 & 0.1625 \\
         \textit{w/} single-view & 0.2886 & 0.4120 & 0.1286 & 0.0964 & 0.2936 & 0.3381 & 0.1498 & 0.2482 & 0.1694 & 0.1386 & 0.1816 & 0.2086 \\
         \textit{w/o} gate & 0.2895 & 0.4167 & 0.1284 & 0.0972 & 0.2947 & 0.3418 & 0.1523& 0.2455& 0.1699& 0.1373& 0.1816& 0.2066\\
         \textit{w/o} $\mathcal{L}_{\mathrm{ortho}}$ & 0.2921 & 0.4141 & 0.1307 & 0.0971 & 0.2996 & 0.3405 & 0.1420 & 0.2321 & 0.1617 & 0.1313 & 0.1733 & 0.1973 \\
         \textit{w/o} $\mathcal{L}_{\mathrm{lb}}$ & 0.2921 & 0.4153 & 0.1300 & 0.0972 & 0.2981 & 0.3414 & 0.1436 & 0.2348 & 0.1631 & 0.1329 & 0.1765 & 0.2007 \\
         \textit{w/o} $\mathcal{L}_{\mathrm{lb}}, \mathcal{L}_{\mathrm{ortho}}$ & 0.2924 & 0.4147 & 0.1307 & 0.0972 & \textbf{0.3001} & 0.3412 & 0.1453 & 0.2405 & 0.1652 & 0.1354 & 0.1787 & 0.2044 \\
    
     \bottomrule
    \end{tabular}
    }
\end{table*}

\subsection{Ablation Study (RQ2)}
In this section, we conduct an ablation study on CiteULike and XING-5000 to evaluate the contribution of each component in PMFRec. The results are summarized in Table~\ref{tab:ablation_full}. We compare the full model against several variants: \textit{w/o} $P_u$ (using only the global component $P_g$), \textit{w/} only $P_u$ (using only the personalized component), \textit{w/} single-view ($K=1$), \textit{w/o} gate (uniform averaging of views), and versions excluding regularization terms ($\mathcal{L}_{\mathrm{ortho}}$, $\mathcal{L}_{\mathrm{lb}}$).

The most significant performance degradation occurs when personalization is entirely removed (\textit{w/o} $P_u$) or when the model relies solely on personalized representations (\textit{w/} only $P_u$). The poor performance of \textit{w/} only $P_u$ stems from the inability to leverage collaborative knowledge shared across other users via the global item representation. Conversely, the drop in \textit{w/o} $P_u$ validates our core premise that integrating user-specific preference signals is essential for effective cold-start recommendation. These results confirm that PMFRec successfully captures robust user preferences by fusing personalized and global perspectives into a single-type item embedding.

Beyond personalization, the architectural choices for the global encoder significantly impact performance. Utilizing a single-view encoder (\textit{w/} single-view) hurts accuracy, supporting the necessity of multi-view encoding to resolve compositionality failures and capture conflicting semantic facets. Furthermore, removing the gating module (\textit{w/o} gate), which results in simple uniform averaging, leads to a performance drop. This indicates the importance of item-adaptive view selection, where each item can selectively emphasize its most relevant semantic subspaces through our gating mechanism.

Finally, excluding the orthogonality loss ($\mathcal{L}_{\mathrm{ortho}}$) or the load-balancing loss ($\mathcal{L}_{\mathrm{lb}}$) consistently leads to suboptimal results. Interestingly, as shown in Table~\ref{tab:ablation_full}, removing both terms simultaneously can sometimes yield better results than removing only one (e.g., on CiteULike). This suggests a strong inter-dependency between the two objectives, implying they are most effective when working in tandem to encourage diverse and balanced view utilization across the item pool.


\begin{table}
\caption{Comparison of zero-score ratio (Z@K).}
    \centering
    \small
   \resizebox{1.0\linewidth}{!}{%
        \begin{tabular}{cc|ccccc}
        \toprule
           Dataset & Metric  & IFedNCF  & IPFedRec  & \textit{w/o} $P_u$ & w/ single-view & PMFRec\\
             \midrule
             \multirow{2}{*}{CiteULike} & Z@10 & 36.93\% &  37.15\%  & 37.65\%  & 34.89\% & \textbf{34.71}\%\\
                                        & Z@20 & 24.23\% &  25.73\%  & 25.24\%  & 23.66\% & \textbf{22.92}\%\\
            \midrule
             \multirow{2}{*}{XING-5000} & Z@10 & 30.30\% &  32.00\%  & 30.34\% & 28.94\% & \textbf{26.80}\%\\
                                        & Z@20 & 12.12\% &  11.24\% &11.86\%  & 11.46\% & \textbf{11.04}\% \\
         \bottomrule
        \end{tabular}
    }

    \label{tab:zero_score}
\end{table}

\begin{figure}
    \centering
    \includegraphics[width=1.0\linewidth]{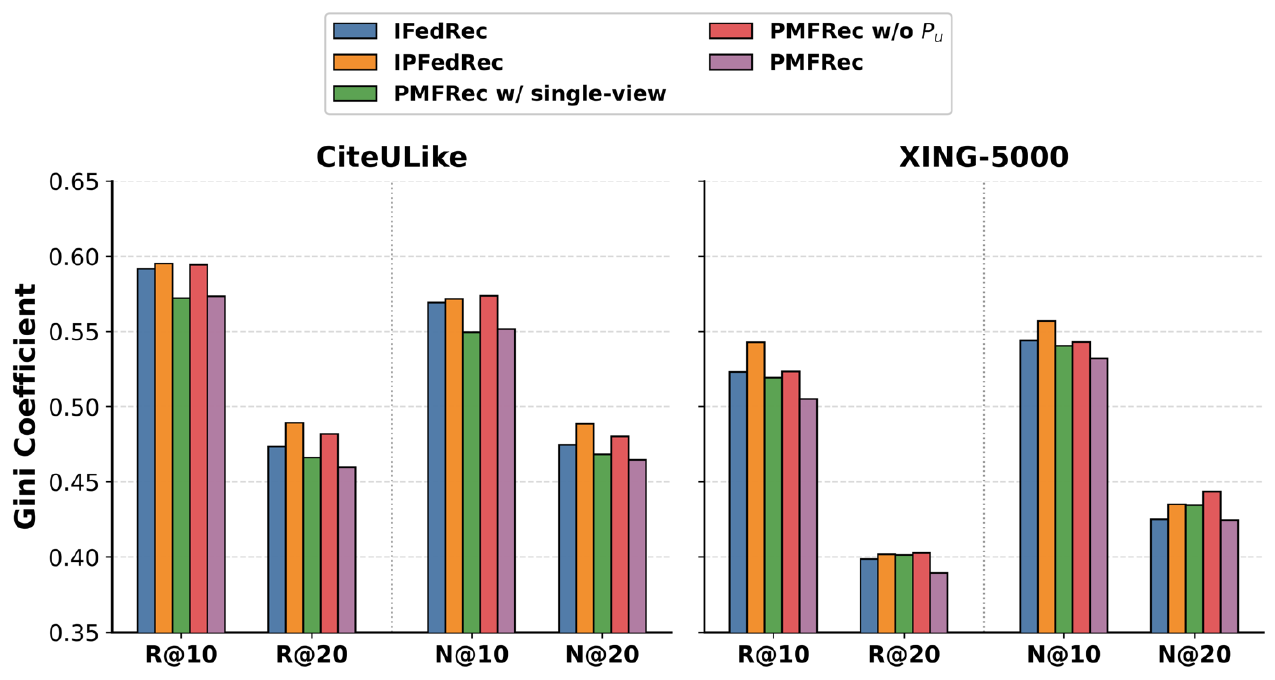}
    \caption{Analysis for user-level fairness and performance distribution.}
    \label{fig:gini}
\end{figure}

\subsection{Personalization Analysis (RQ3)}
In this section, we analyze how our \emph{personalized item representations} affect cold-start behavior on CiteULike and XING-5000.
Because prior methods rely on aggregation and attribute-feature-to-embedding alignment, they can \emph{dilute} user-specific preference signals, leading to score shrinkage and overly conservative rankings. To quantify this effect, we report the \emph{zero-score ratio} (Z@$K$) in the cold setting, defined as the fraction of users whose cold-item Recall@$K$ is zero. Table~\ref{tab:zero_score} shows that PMFRec consistently reduces the zero-score ratio compared to the baselines, whereas removing personalization increases it. This verifies that personalized representations effectively mitigate preference dilution and help more users receive at least one correct cold-item recommendation.

We further assess user-level fairness using the Gini coefficient, where lower values indicate more evenly distributed per-user performance\footnote{Let $n$ be the number of users and let $x_i \ge 0$ denote the per-user metric value
(e.g., Recall@$K$ or NDCG@$K$) for user $i$.
Let $x_{(1)} \le \cdots \le x_{(n)}$ be the sorted values, and assume $\sum_{i=1}^{n} x_{(i)} > 0$.
We compute the Gini coefficient as $G= \frac{\sum_{i=1}^{n} (2i-n-1)\,x_{(i)}}{n \sum_{i=1}^{n} x_{(i)}}$.
Lower $G$ indicates more evenly distributed performance across users.}. As shown in Figure~\ref{fig:gini}, PMFRec generally achieves lower Gini coefficients than the baselines, while removing personalization increases inequality across users.

Notably, personalized representations are the primary factor driving these gains. Even when the multi-view component is not used, the single-view personalized variant already improves both zero-score ratio and user-level fairness over non-personalized alternatives. Although combining personalization with multi-view representation yields the best overall performance in most cases, it is not uniformly superior on every fairness metric; for example, on CiteULike, the full model shows a slight increase in Gini for R@10 and N@10 compared with the single-view personalized variant. Nevertheless, both personalized variants consistently outperform the corresponding non-personalized settings, confirming that personalization is the key mechanism for alleviating preference dilution, while multi-view representation provides complementary gains by modeling more complex relationships among heterogeneous attributes.

Taken together, these results show that PMFRec not only improves overall cold-start accuracy, but also reduces zero-score users and distributes performance more broadly across users by generating more expressive and user-aligned cold-item representations.

\begin{figure}[t]
    \centering
    \includegraphics[width=1.0\linewidth]{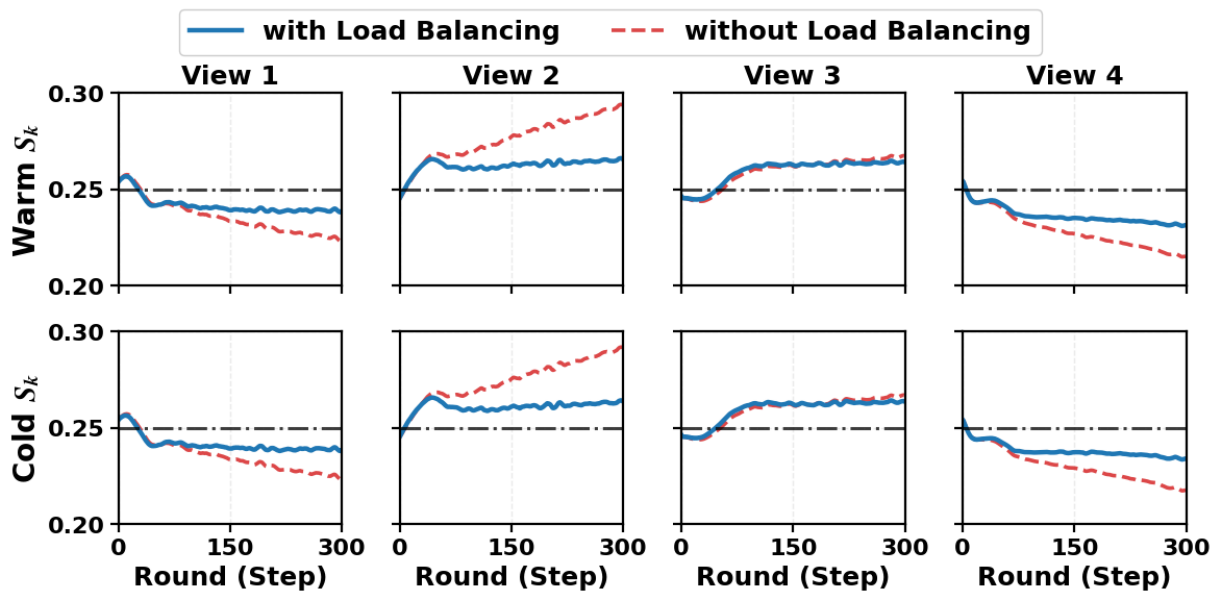}
    \caption{Effect of load balancing loss on CiteULike.}
    \label{fig:load_balancing}
\end{figure}

\begin{figure}[t]
    \centering
    \includegraphics[width=1.0\linewidth]{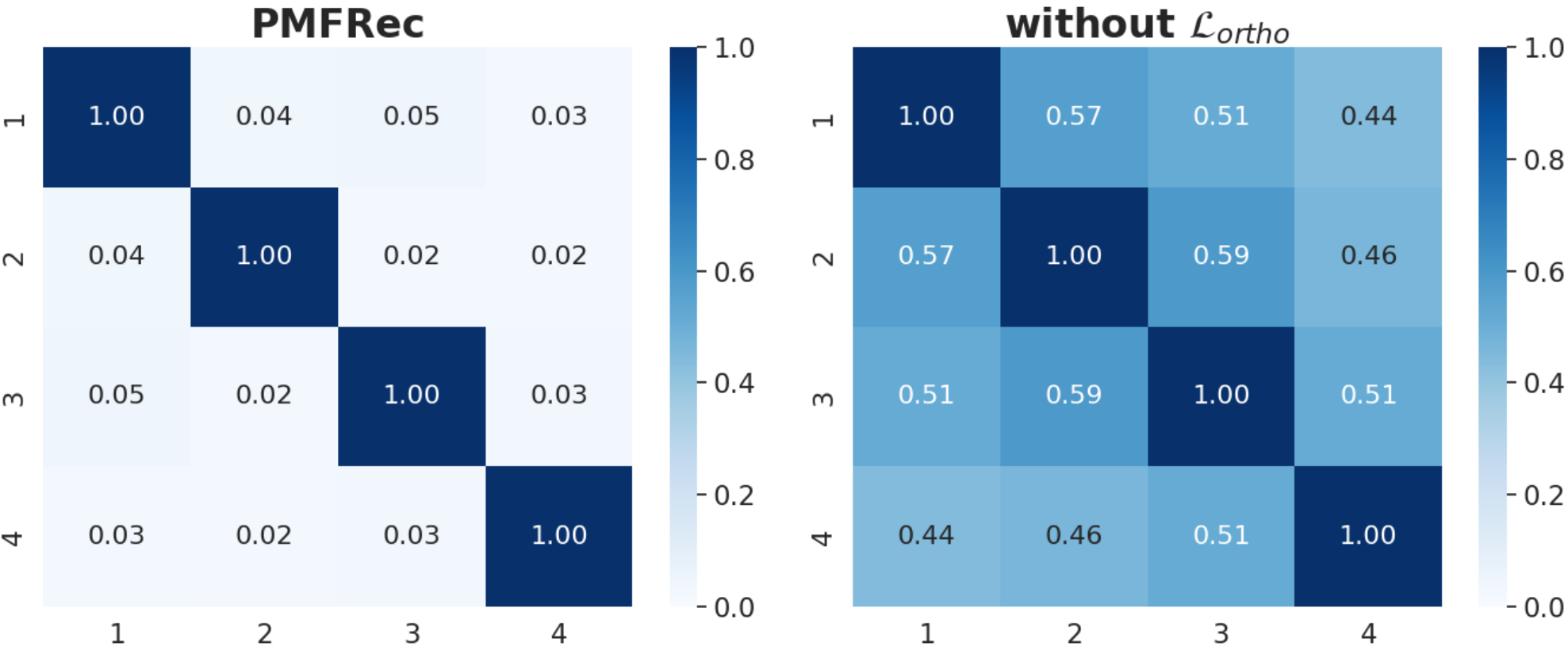}
    \caption{Analysis of cross-view redundancy on CiteULike.}
    \label{fig:orthogonality}
\end{figure}

\begin{table}[t]
\caption{Qualitative Case Study on CiteULike. The items assigned highest weights for different views reveal distinct semantic clusters (e.g., Biology vs. Social Science), supporting the interpretability of our multi-view decomposition.}
    \centering
    \resizebox{1.0\linewidth}{!}{%
    \begin{tabular}{c|l}
    \toprule
        View & Title of literature \\
        \midrule
        \multirow{3}{*}{View 1} &   Evolving protein interaction networks through gene duplication  \\
         &  Light Microscopy Techniques for Live Cell Imaging  \\
         &  Widespread occurrence of antisense transcription in the human genome  \\
     \midrule
        \multirow{3}{*}{View 2}&  Thirteen Ways to Look at the Correlation Coefficient \\
         & Information Literacy  \\
         &   A face(book) in the crowd: social Searching vs. social browsing \\
     \bottomrule
    \end{tabular}
    }
    \label{tab:case_study}
\end{table}

\subsection{Multi-view Analysis (RQ4)}
In this section, we analyze the behavior of our server-side multi-view encoder and examine whether the proposed design indeed produces diverse and meaningful view decomposition.

We first investigate whether $\mathcal{L}_{\mathrm{ortho}}$ encourages distinct view representations.
Figure~\ref{fig:orthogonality} reports the average pairwise cosine similarity between view-specific projected representations on CiteULike, computed from \emph{unseen cold-item attribute features}.
With $\mathcal{L}_{\mathrm{ortho}}$, the cross-view similarities become substantially lower than those obtained without it, indicating that different views capture less redundant and more complementary semantics.
In contrast, without the orthogonality regularization, the projected representations from different views remain highly correlated, suggesting that multiple views collapse into overlapping feature subspaces rather than providing genuinely distinct perspectives.

Next, we study whether load balancing prevents routing collapse.
Figure~\ref{fig:load_balancing} plots the routing masses $S_k$ over training steps for both warm and cold items.
With load balancing, the routing mass assigned to each view remains more even and stable throughout training, whereas without it, routing progressively becomes skewed toward a subset of views.
This pattern indicates that the load balancing objective is crucial for preventing dominant-view collapse and for ensuring that all views participate in representing items, including unseen cold items.

Finally, Table~\ref{tab:case_study} presents a qualitative case study.
For two representative views, we list the items assigned the largest gating weights.
The resulting groups exhibit clearly different semantics, such as biology-oriented versus information-science-oriented topics, suggesting that the learned views are not only computationally distinct but also semantically interpretable.
Taken together, these results show that the proposed multi-view encoder learns complementary, non-collapsed, and interpretable feature subspaces, which helps explain why the full model consistently outperforms its single-view counterpart.

\begin{figure}[t]
    \centering
    \includegraphics[width=\linewidth]{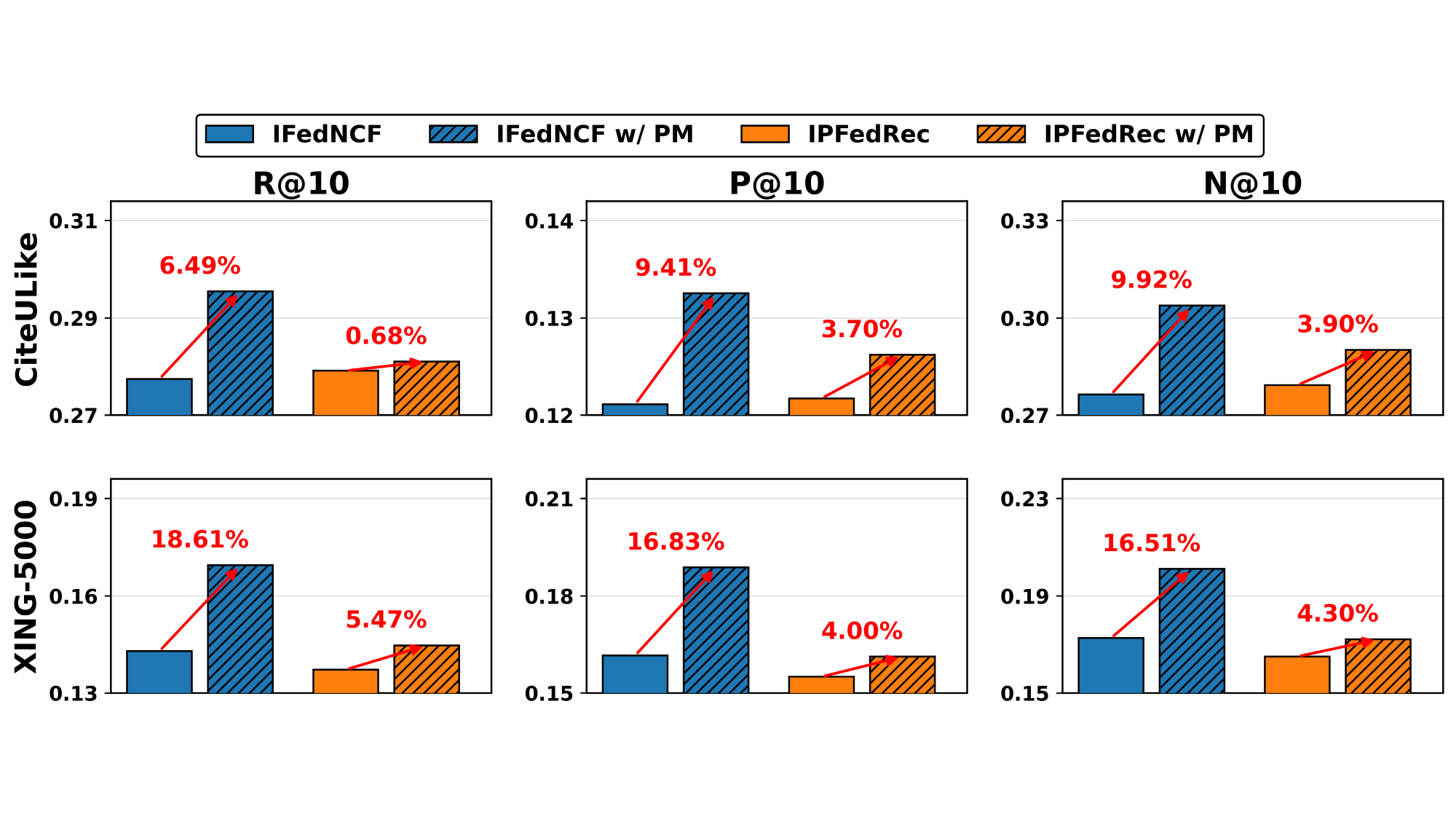}
    \caption{Adaptability of personalized and multi-view representations to existing baselines.}
    \label{fig:baseline_pm}
\end{figure}

\begin{figure}[t]
    \centering
    \includegraphics[width=\linewidth]{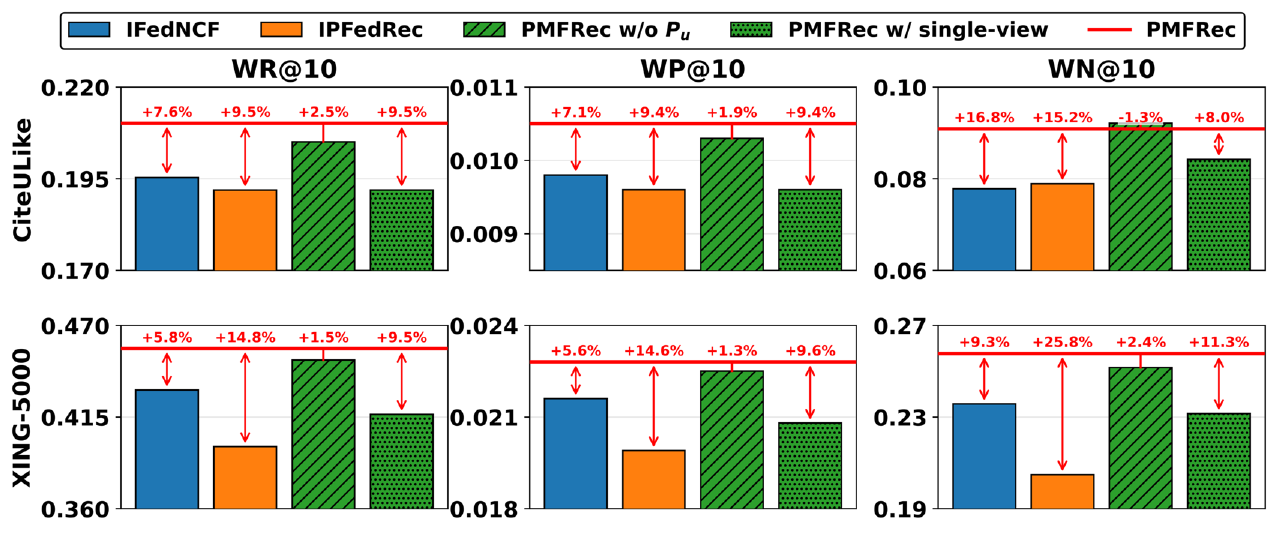}
    \caption{Performance in the warm scenario.}
    \label{fig:warm}
\end{figure}

\subsection{Adaptability and Warm Scenario (RQ5)}
In this section, we evaluate whether our proposed personalized and multi-view representations can be seamlessly integrated into existing baselines (e.g., IFedNCF, IPFedRec), and assess the overall effectiveness of our method in a warm recommendation scenario. For clarity, the suffix \textit{w/} PM denotes a variant where the baseline's original item representations are replaced with our \textbf{P}ersonalized and \textbf{M}ulti-view representations, while the rest of the baseline architecture remains unchanged.

As illustrated in Figure~\ref{fig:baseline_pm}, integrating our representations into prior frameworks yields substantial performance gains, ranging from 0.68\% to 18.61\%. This demonstrates the high adaptability of our proposed representations. However, it is worth noting that these augmented baselines must transmit the two distinct types of representations separately to the clients, which can introduce communication inefficiencies during the training process.

Furthermore, Figure~\ref{fig:warm} demonstrates that our proposed method consistently achieves outstanding performance even in the warm scenario. Overall, our method significantly outperforms the representative baselines, achieving up to a 25.8\% performance improvement. 
We also investigate which component—the personalized representation or the multi-view encoding—had a more significant impact on performance in the warm scenario. Interestingly, in contrast to the cold scenario, the multi-view knowledge proved to be the more critical factor. Unlike in the cold setting, removing the multi-view encoding resulted in a substantial performance drop in the warm setting. This suggests that while personalization is highly effective for addressing unseen item features (cold items), multi-view knowledge becomes paramount when abundant training signals are available. Although personalization remains beneficial, the ability to comprehensively interpret the features of warm item embeddings through multiple views is essential to maximize performance amid rich training signals.



\begin{figure}[h]
    \centering
    \includegraphics[width=1.0\linewidth]{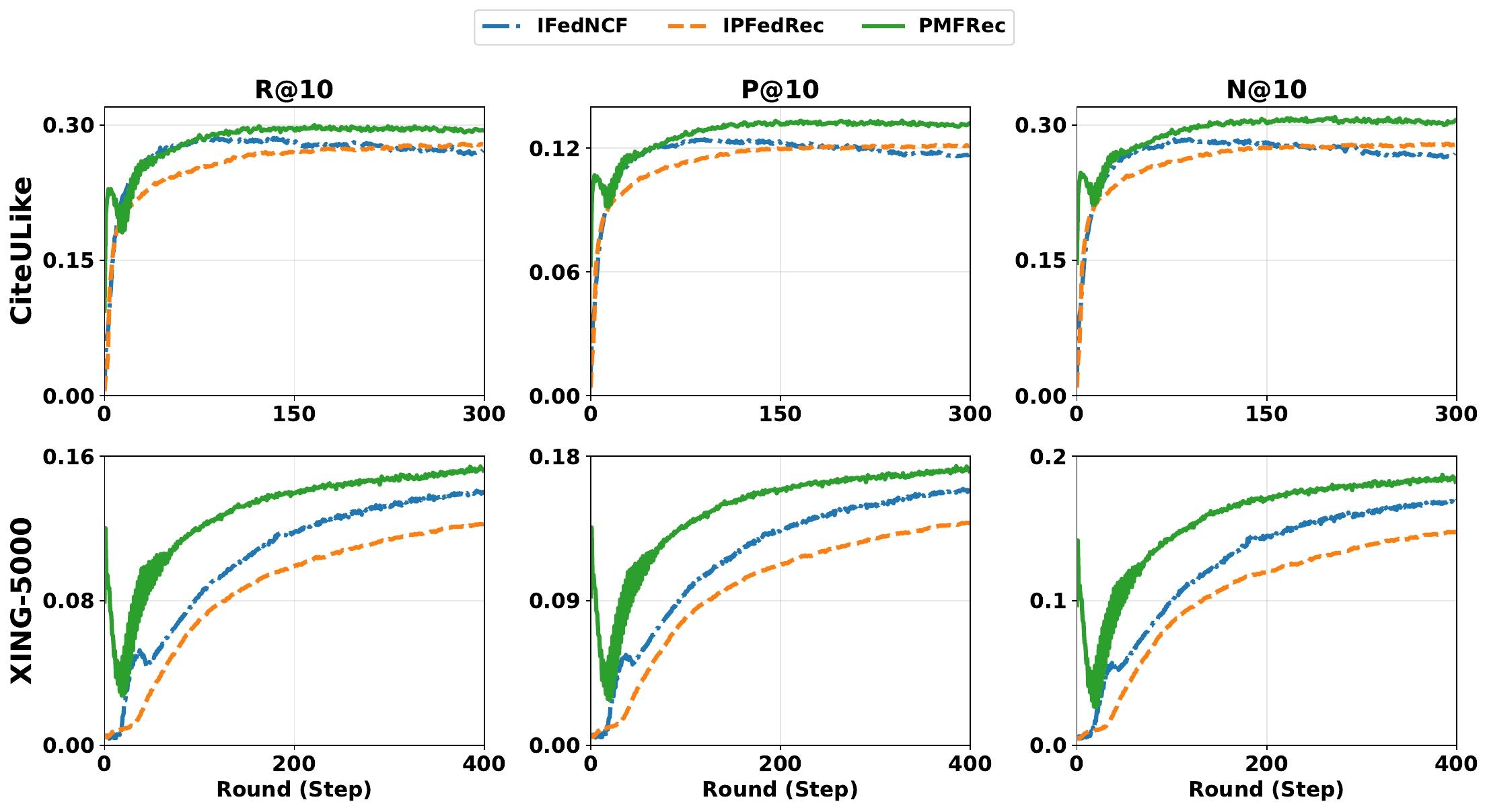}
    \caption{Convergence curves on CiteULike.}
    \label{fig:converge}
\end{figure}

\begin{figure} [h]
    \centering
    \includegraphics[width=1.0\linewidth]{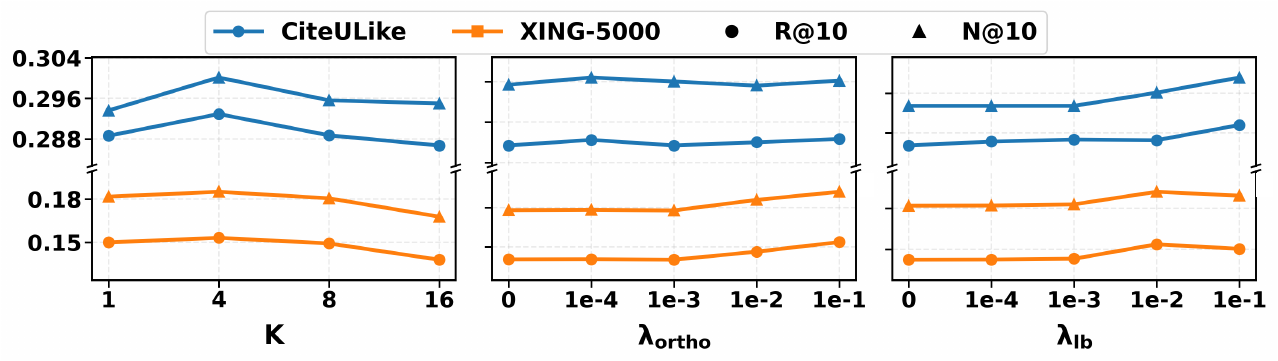}
    \caption{Parameter sensitivity analysis on CiteULike and XING-5000.}
    \label{fig:sensitivity}
\end{figure}

\subsection{Convergence and Hyperparameter Analysis (RQ6)}
\label{sec:rq7}
\noindent \textbf{Convergence Analysis.} Figure~\ref{fig:converge} shows stable convergence on CiteULike and XING-5000, and we observe similar trends on the other datasets. Unlike the other baselines, PMFRec exhibits a brief performance drop in the early stage followed by a steady improvement. We attribute this to applying personalization before user preferences are sufficiently formed. In contrast, most baselines train remain relatively stable and performance increases more monotonically. Nevertheless, the final performance of \textsc{PMFRec} is higher overall.

\noindent \textbf{Parameter Sensitivity.} Figure~\ref{fig:sensitivity} studies the sensitivity to $K$, $\lambda_{\mathrm{ortho}}$, and $\lambda_{\mathrm{lb}}$ on both datasets. For the number of views, $K=4$ yields the best performance on both CiteULike and XING-5000, and performance generally decreases as $K$ grows further. For $\lambda_{\mathrm{ortho}}$, the best value is $10^{-4}$ on CiteULike and $10^{-1}$ on XING-5000; nevertheless, performance is consistently better with a positive $\lambda_{\mathrm{ortho}}$ than with $\lambda_{\mathrm{ortho}}=0$, and the overall variation is modest across a wide range. For $\lambda_{\mathrm{lb}}$, performance improves as $\lambda_{\mathrm{lb}}$ increases on both datasets, suggesting that more balanced view utilization (i.e., fairer routing across views) is beneficial; again, the sensitivity is not severe.

\begin{table}[t]
    \caption{Performance under (ε, δ)-LDP with varying privacy budgets ε and fixed $\delta$.}
    \centering
    \resizebox{1.0\linewidth}{!}{%
        \begin{tabular}{cl|cc|cc|cc}
            \toprule
            \multirow{2}{*}{Dataset} & Method & \multicolumn{2}{c|}{IFedNCF} & \multicolumn{2}{c|}{IFedNCF w/ PM} &  \multicolumn{2}{c}{PMFRec}   \\
             & Metrics & R@10 & N@10 & R@10 & N@10 & R@10 & N@10 \\
             \midrule
             
            \multirow{4}{*}{CiteULike} & \textit{w/o} LDP & 0.2774 & 0.2763 & 0.2954 & 0.3037 & 0.2929 & 0.3001 \\
                                     & $\varepsilon = 4$ & 0.0928& 0.1029&  0.1088&  0.1147& 0.2274& 0.2424\\
                                     & $\varepsilon = 8$ & 0.1556& 0.1640&  0.1734&  0.1776& 0.2381& 0.2494\\
                                     & $\varepsilon = 10$ & 0.1785& 0.1892&  0.1743&  0.1781& 0.2432& 0.2531\\
             \midrule
            \multirow{4}{*}{XING-5000} & \textit{w/o} LDP & 0.1429 & 0.1726 & 0.1695 & 0.2011 & 0.1530 & 0.1850 \\
             & $\varepsilon = 4$ &  0.0410&  0.0492& 0.0397&  0.0497&  0.0858&  0.1039\\
             & $\varepsilon = 8$ &  0.0538&  0.0674&  0.0717&  0.0906&  0.0963&  0.1149\\
             & $\varepsilon = 10$ &  0.0585&  0.0739&  0.0833&  0.1042&  0.0992& 0.1193\\
            \bottomrule
        \end{tabular}%
    }
    \label{tab:ldp}
\end{table}

\begin{figure}
    \centering
    \includegraphics[width=1.0\linewidth]{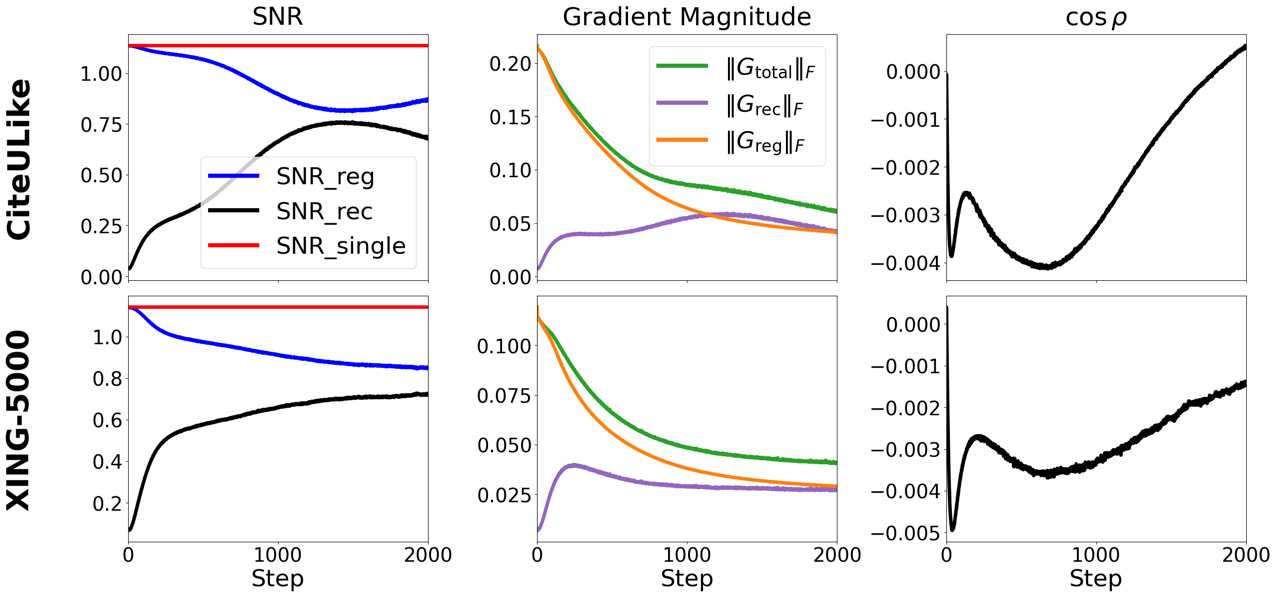}
    \caption{Qualitative case study under LDP ($\varepsilon=10$).
    Per-round diagnostics on CiteULike (top) and XING-5000 (bottom) for IFedNCF w/PM.}
    \label{fig:snr}
\end{figure}

\subsection{Privacy-Preserving Analysis (RQ7)}
In this section, we evaluate the robustness of PMFRec under $(\varepsilon, \delta)$-LDP by varying the privacy budget $\varepsilon$. As shown in Table~\ref{tab:ldp}, while the performance of all models inevitably decreases as $\varepsilon$ becomes smaller due to the increased noise scale, PMFRec exhibits significantly higher robustness compared to regularizer-based baselines. Notably, baselines that perform well in non-private settings suffer a sharp decline under LDP, highlighting a critical privacy-utility trade-off inherent in existing methods.

To investigate the underlying cause of this robustness gap, we analyze the effective projection Signal-to-Noise Ratio (SNR) of the local updates. The SNR along a target direction $\hat{v}$ is defined as follows:
$\mathrm{SNR}_{\hat{v}}^{(+)}(G) = \frac{\min(\|G\|_F, C) \max(\cos\phi_v, 0)}{\sigma C},$
\footnote{Let $\phi_v := \angle(G, \hat{v})$. Based on the noise $Z$ defined in Section~\ref{sec:ldp}, the directional projection of $\hat{v}$ onto the clipped gradient $G$ is given by $\langle \mathrm{clip}_{C}(G) + Z, \hat{v} \rangle$. Therefore, this directional projection follows a Gaussian distribution: $\langle \mathrm{clip}_{C}(G) + Z, \hat{v} \rangle \sim \mathcal{N}\big(\min(\|G\|_F, C)\cos\phi_v, \sigma^2 C^2\big)$. Consequently, the effective projection signal-to-noise ratio is formulated as $\mathrm{SNR}_{\hat{v}}^{(+)}(G) = \frac{\min(\|G\|_F, C) \max(\cos\phi_v, 0)}{\sigma C}$.}
where $G$ is the gradient, $C$ is the clipping threshold, $\sigma$ is the noise scale, and $\phi_v$ is the angle between the gradient and the target direction.

The empirical diagnostics provided in Figure~\ref{fig:snr} reveal the structural limitations of existing schemes. To rigorously analyze these mechanics, we specifically focus on IFedNCF \textit{w/} PM in this figure, as it exhibited the highest baseline performance under LDP, allowing us to dissect the underlying reasons behind its behavior. Specifically, both IFedNCF and IPFedRec rely on an explicit client-side regularizer to align item embeddings with attribute-based representations. Consequently, the local gradient $G_u$ is a composite of two objectives: $G_u = G_{\mathrm{rec}} + G_{\mathrm{reg}}$. This multi-objective formulation leads to a severe \textsc{signal interference} effect. First, we observe that the SNR values for both the recommendation signal ($G_{\mathrm{rec}}$) and the regularization signal ($G_{\mathrm{reg}}$) remain strictly below the single-objective upper bound ($\mathrm{SNR}_{\mathrm{single}}$) throughout training. Second, as the magnitude of the regularizer grows, the effective SNR for the primary recommendation task is heavily suppressed. Furthermore, the alignment term $\cos\rho$ (representing the angle between $G_{\mathrm{rec}}$ and $G_{\mathrm{reg}}$) is negative for a substantial portion of the training process, indicating that these two gradients frequently point in opposing directions. 

When clipped and injected with LDP noise, this gradient misalignment exacerbates directional distortion, as the true signal is drowned out by the conflicting objectives, preventing the model from reaching the desired optimum. PMFRec fundamentally avoids this issue. By eliminating the client-side regularizer, PMFRec maintains a singular, clean gradient signal for recommendation, resulting in stable optimization and superior robustness even in high-noise regimes.

\section{Conclusion}
In this paper, we proposed PMFRec, a personalized and multi-view federated framework for cold-start recommendation under dual-sided constraints. PMFRec generates user-specific representations for cold items by combining a personalized encoder with a global multi-view attribute encoder equipped with item-adaptive gating. By modeling heterogeneous attribute semantics through complementary views and selectively composing them for each item, our framework alleviates the representation averaging and semantic entanglement problems of prior single-mapping approaches. In addition, PMFRec integrates collaborative and attribute knowledge into a single exchanged item representation, eliminating the need for an explicit client-side regularizer and improving communication efficiency during training. Extensive experiments showed that PMFRec consistently outperforms strong baselines, while ablation studies and analyses confirmed the importance of personalization, multi-view encoding, adaptive gating, and view regularization. Furthermore, PMFRec improves user-level fairness, transfers effectively to warm-scenario settings, and remains more robust than regularizer-based approaches under Local Differential Privacy (LDP). Overall, these results demonstrate that PMFRec provides an effective and practical solution for federated cold-start recommendation in both standard and privacy-preserving environments.

\bibliographystyle{ACM-Reference-Format}
\bibliography{0_ref}


\begin{thebibliography}{38}


\ifx \showCODEN    \undefined \def \showCODEN     #1{\unskip}     \fi
\ifx \showISBNx    \undefined \def \showISBNx     #1{\unskip}     \fi
\ifx \showISBNxiii \undefined \def \showISBNxiii  #1{\unskip}     \fi
\ifx \showISSN     \undefined \def \showISSN      #1{\unskip}     \fi
\ifx \showLCCN     \undefined \def \showLCCN      #1{\unskip}     \fi
\ifx \shownote     \undefined \def \shownote      #1{#1}          \fi
\ifx \showarticletitle \undefined \def \showarticletitle #1{#1}   \fi
\ifx \showURL      \undefined \def \showURL       {\relax}        \fi
\providecommand\bibfield[2]{#2}
\providecommand\bibinfo[2]{#2}
\providecommand\natexlab[1]{#1}
\providecommand\showeprint[2][]{arXiv:#2}

\bibitem[Abadi et~al\mbox{.}(2016)]%
        {ldp}
\bibfield{author}{\bibinfo{person}{Martin Abadi}, \bibinfo{person}{Andy Chu}, \bibinfo{person}{Ian Goodfellow}, \bibinfo{person}{H~Brendan McMahan}, \bibinfo{person}{Ilya Mironov}, \bibinfo{person}{Kunal Talwar}, {and} \bibinfo{person}{Li Zhang}.} \bibinfo{year}{2016}\natexlab{}.
\newblock \showarticletitle{Deep learning with differential privacy}. In \bibinfo{booktitle}{\emph{Proceedings of the 2016 ACM SIGSAC conference on computer and communications security}}. \bibinfo{pages}{308--318}.
\newblock
\href{https://doi.org/10.1145/2976749.2978318}{doi:\nolinkurl{10.1145/2976749.2978318}}


\bibitem[Abel et~al\mbox{.}(2017)]%
        {xing}
\bibfield{author}{\bibinfo{person}{Fabian Abel}, \bibinfo{person}{Yashar Deldjoo}, \bibinfo{person}{Mehdi Elahi}, {and} \bibinfo{person}{Daniel Kohlsdorf}.} \bibinfo{year}{2017}\natexlab{}.
\newblock \showarticletitle{Recsys challenge 2017: Offline and online evaluation}. In \bibinfo{booktitle}{\emph{Proceedings of the eleventh acm conference on recommender systems}}. \bibinfo{pages}{372--373}.
\newblock
\href{https://doi.org/10.1145/3109859.3109954}{doi:\nolinkurl{10.1145/3109859.3109954}}


\bibitem[Ammad-Ud-Din et~al\mbox{.}(2019)]%
        {fcf}
\bibfield{author}{\bibinfo{person}{Muhammad Ammad-Ud-Din}, \bibinfo{person}{Elena Ivannikova}, \bibinfo{person}{Suleiman~A Khan}, \bibinfo{person}{Were Oyomno}, \bibinfo{person}{Qiang Fu}, \bibinfo{person}{Kuan~Eeik Tan}, {and} \bibinfo{person}{Adrian Flanagan}.} \bibinfo{year}{2019}\natexlab{}.
\newblock \showarticletitle{Federated collaborative filtering for privacy-preserving personalized recommendation system}.
\newblock \bibinfo{journal}{\emph{arXiv preprint arXiv:1901.09888}} (\bibinfo{year}{2019}).
\newblock
\href{https://doi.org/10.48550/arXiv.1901.09888}{doi:\nolinkurl{10.48550/arXiv.1901.09888}}


\bibitem[Anwaar et~al\mbox{.}(2018)]%
        {hybrid_1}
\bibfield{author}{\bibinfo{person}{Fahad Anwaar}, \bibinfo{person}{Naima Iltaf}, \bibinfo{person}{Hammad Afzal}, {and} \bibinfo{person}{Raheel Nawaz}.} \bibinfo{year}{2018}\natexlab{}.
\newblock \showarticletitle{HRS-CE: A hybrid framework to integrate content embeddings in recommender systems for cold start items}.
\newblock \bibinfo{journal}{\emph{Journal of computational science}}  \bibinfo{volume}{29} (\bibinfo{year}{2018}), \bibinfo{pages}{9--18}.
\newblock
\href{https://doi.org/10.1016/j.jocs.2018.09.008}{doi:\nolinkurl{10.1016/j.jocs.2018.09.008}}


\bibitem[Bai et~al\mbox{.}(2023)]%
        {gorec}
\bibfield{author}{\bibinfo{person}{Haoyue Bai}, \bibinfo{person}{Min Hou}, \bibinfo{person}{Le Wu}, \bibinfo{person}{Yonghui Yang}, \bibinfo{person}{Kun Zhang}, \bibinfo{person}{Richang Hong}, {and} \bibinfo{person}{Meng Wang}.} \bibinfo{year}{2023}\natexlab{}.
\newblock \showarticletitle{Gorec: a generative cold-start recommendation framework}. In \bibinfo{booktitle}{\emph{Proceedings of the 31st ACM international conference on multimedia}}. \bibinfo{pages}{1004--1012}.
\newblock
\href{https://doi.org/10.1145/3581783.3612238}{doi:\nolinkurl{10.1145/3581783.3612238}}


\bibitem[Chai et~al\mbox{.}(2020)]%
        {fedmf}
\bibfield{author}{\bibinfo{person}{Di Chai}, \bibinfo{person}{Leye Wang}, \bibinfo{person}{Kai Chen}, {and} \bibinfo{person}{Qiang Yang}.} \bibinfo{year}{2020}\natexlab{}.
\newblock \showarticletitle{Secure federated matrix factorization}.
\newblock \bibinfo{journal}{\emph{IEEE Intelligent Systems}} \bibinfo{volume}{36}, \bibinfo{number}{5} (\bibinfo{year}{2020}), \bibinfo{pages}{11--20}.
\newblock
\href{https://doi.org/10.1109/MIS.2020.3014880}{doi:\nolinkurl{10.1109/MIS.2020.3014880}}


\bibitem[Chen et~al\mbox{.}(2022)]%
        {gar}
\bibfield{author}{\bibinfo{person}{Hao Chen}, \bibinfo{person}{Zefan Wang}, \bibinfo{person}{Feiran Huang}, \bibinfo{person}{Xiao Huang}, \bibinfo{person}{Yue Xu}, \bibinfo{person}{Yishi Lin}, \bibinfo{person}{Peng He}, {and} \bibinfo{person}{Zhoujun Li}.} \bibinfo{year}{2022}\natexlab{}.
\newblock \showarticletitle{Generative adversarial framework for cold-start item recommendation}. In \bibinfo{booktitle}{\emph{Proceedings of the 45th International ACM SIGIR Conference on Research and Development in Information Retrieval}}. \bibinfo{pages}{2565--2571}.
\newblock
\href{https://doi.org/10.1145/3477495.3531897}{doi:\nolinkurl{10.1145/3477495.3531897}}


\bibitem[Cheng et~al\mbox{.}(2016)]%
        {wdr}
\bibfield{author}{\bibinfo{person}{Heng-Tze Cheng}, \bibinfo{person}{Levent Koc}, \bibinfo{person}{Jeremiah Harmsen}, \bibinfo{person}{Tal Shaked}, \bibinfo{person}{Tushar Chandra}, \bibinfo{person}{Hrishi Aradhye}, \bibinfo{person}{Glen Anderson}, \bibinfo{person}{Greg Corrado}, \bibinfo{person}{Wei Chai}, \bibinfo{person}{Mustafa Ispir}, {et~al\mbox{.}}} \bibinfo{year}{2016}\natexlab{}.
\newblock \showarticletitle{Wide \& deep learning for recommender systems}. In \bibinfo{booktitle}{\emph{Proceedings of the 1st workshop on deep learning for recommender systems}}. \bibinfo{pages}{7--10}.
\newblock
\href{https://doi.org/10.1145/2988450.2988454}{doi:\nolinkurl{10.1145/2988450.2988454}}


\bibitem[Dwork et~al\mbox{.}(2014)]%
        {ldp_125}
\bibfield{author}{\bibinfo{person}{Cynthia Dwork}, \bibinfo{person}{Aaron Roth}, {et~al\mbox{.}}} \bibinfo{year}{2014}\natexlab{}.
\newblock \showarticletitle{The algorithmic foundations of differential privacy}.
\newblock \bibinfo{journal}{\emph{Foundations and trends{\textregistered} in theoretical computer science}} \bibinfo{volume}{9}, \bibinfo{number}{3--4} (\bibinfo{year}{2014}), \bibinfo{pages}{211--407}.
\newblock
\href{https://doi.org/10.1561/0400000042}{doi:\nolinkurl{10.1561/0400000042}}


\bibitem[Flanagan et~al\mbox{.}(2020)]%
        {fedmvmf}
\bibfield{author}{\bibinfo{person}{Adrian Flanagan}, \bibinfo{person}{Were Oyomno}, \bibinfo{person}{Alexander Grigorievskiy}, \bibinfo{person}{Kuan~E Tan}, \bibinfo{person}{Suleiman~A Khan}, {and} \bibinfo{person}{Muhammad Ammad-Ud-Din}.} \bibinfo{year}{2020}\natexlab{}.
\newblock \showarticletitle{Federated multi-view matrix factorization for personalized recommendations}. In \bibinfo{booktitle}{\emph{Joint European conference on machine learning and knowledge discovery in databases}}. Springer, \bibinfo{pages}{324--347}.
\newblock
\href{https://doi.org/10.1007/978-3-030-67661-2_20}{doi:\nolinkurl{10.1007/978-3-030-67661-2_20}}


\bibitem[Han et~al\mbox{.}(2025)]%
        {fedcia}
\bibfield{author}{\bibinfo{person}{Mingzhe Han}, \bibinfo{person}{Dongsheng Li}, \bibinfo{person}{Jiafeng Xia}, \bibinfo{person}{Jiahao Liu}, \bibinfo{person}{Hansu Gu}, \bibinfo{person}{Peng Zhang}, \bibinfo{person}{Ning Gu}, {and} \bibinfo{person}{Tun Lu}.} \bibinfo{year}{2025}\natexlab{}.
\newblock \showarticletitle{FedCIA: Federated collaborative information aggregation for privacy-preserving recommendation}. In \bibinfo{booktitle}{\emph{Proceedings of the 48th International ACM SIGIR Conference on Research and Development in Information Retrieval}}. \bibinfo{pages}{1687--1696}.
\newblock
\href{https://doi.org/10.1145/3726302.3729977}{doi:\nolinkurl{10.1145/3726302.3729977}}


\bibitem[He and McAuley(2016)]%
        {vbpr}
\bibfield{author}{\bibinfo{person}{Ruining He} {and} \bibinfo{person}{Julian McAuley}.} \bibinfo{year}{2016}\natexlab{}.
\newblock \showarticletitle{VBPR: visual bayesian personalized ranking from implicit feedback}. In \bibinfo{booktitle}{\emph{Proceedings of the AAAI conference on artificial intelligence}}, Vol.~\bibinfo{volume}{30}.
\newblock
\href{https://doi.org/10.1609/aaai.v30i1.9973}{doi:\nolinkurl{10.1609/aaai.v30i1.9973}}


\bibitem[He et~al\mbox{.}(2020)]%
        {lightgcn}
\bibfield{author}{\bibinfo{person}{Xiangnan He}, \bibinfo{person}{Kuan Deng}, \bibinfo{person}{Xiang Wang}, \bibinfo{person}{Yan Li}, \bibinfo{person}{Yongdong Zhang}, {and} \bibinfo{person}{Meng Wang}.} \bibinfo{year}{2020}\natexlab{}.
\newblock \showarticletitle{Lightgcn: Simplifying and powering graph convolution network for recommendation}. In \bibinfo{booktitle}{\emph{Proceedings of the 43rd International ACM SIGIR conference on research and development in Information Retrieval}}. \bibinfo{pages}{639--648}.
\newblock
\href{https://doi.org/10.1145/3397271.3401063}{doi:\nolinkurl{10.1145/3397271.3401063}}


\bibitem[He et~al\mbox{.}(2017)]%
        {ncf}
\bibfield{author}{\bibinfo{person}{Xiangnan He}, \bibinfo{person}{Lizi Liao}, \bibinfo{person}{Hanwang Zhang}, \bibinfo{person}{Liqiang Nie}, \bibinfo{person}{Xia Hu}, {and} \bibinfo{person}{Tat-Seng Chua}.} \bibinfo{year}{2017}\natexlab{}.
\newblock \showarticletitle{Neural collaborative filtering}. In \bibinfo{booktitle}{\emph{Proceedings of the 26th international conference on world wide web}}. \bibinfo{pages}{173--182}.
\newblock
\href{https://doi.org/10.1145/3038912.3052569}{doi:\nolinkurl{10.1145/3038912.3052569}}


\bibitem[He et~al\mbox{.}(2024)]%
        {cofedrec}
\bibfield{author}{\bibinfo{person}{Xinrui He}, \bibinfo{person}{Shuo Liu}, \bibinfo{person}{Jacky Keung}, {and} \bibinfo{person}{Jingrui He}.} \bibinfo{year}{2024}\natexlab{}.
\newblock \showarticletitle{Co-clustering for federated recommender system}. In \bibinfo{booktitle}{\emph{Proceedings of the ACM Web Conference 2024}}. \bibinfo{pages}{3821--3832}.
\newblock
\href{https://doi.org/10.1145/3589334.3645626}{doi:\nolinkurl{10.1145/3589334.3645626}}


\bibitem[Kim et~al\mbox{.}(2024)]%
        {hybrid_2}
\bibfield{author}{\bibinfo{person}{Jinri Kim}, \bibinfo{person}{Eungi Kim}, \bibinfo{person}{Kwangeun Yeo}, \bibinfo{person}{Yujin Jeon}, \bibinfo{person}{Chanwoo Kim}, \bibinfo{person}{Sewon Lee}, {and} \bibinfo{person}{Joonseok Lee}.} \bibinfo{year}{2024}\natexlab{}.
\newblock \showarticletitle{Content-based graph reconstruction for cold-start item recommendation}. In \bibinfo{booktitle}{\emph{Proceedings of the 47th International ACM SIGIR Conference on Research and Development in Information Retrieval}}. \bibinfo{pages}{1263--1273}.
\newblock
\href{https://doi.org/10.1145/3626772.3657801}{doi:\nolinkurl{10.1145/3626772.3657801}}


\bibitem[Li et~al\mbox{.}(2025b)]%
        {fedcolduser}
\bibfield{author}{\bibinfo{person}{Yichen Li}, \bibinfo{person}{Yijing Shan}, \bibinfo{person}{Yi Liu}, \bibinfo{person}{Haozhao Wang}, \bibinfo{person}{Wei Wang}, \bibinfo{person}{Yi Wang}, {and} \bibinfo{person}{Ruixuan Li}.} \bibinfo{year}{2025}\natexlab{b}.
\newblock \showarticletitle{Personalized Federated Recommendation for Cold-Start Users via Adaptive Knowledge Fusion}. In \bibinfo{booktitle}{\emph{Proceedings of the ACM on Web Conference 2025}}. \bibinfo{pages}{2700--2709}.
\newblock
\href{https://doi.org/10.1145/3696410.3714635}{doi:\nolinkurl{10.1145/3696410.3714635}}


\bibitem[Li et~al\mbox{.}(2024)]%
        {fedrap}
\bibfield{author}{\bibinfo{person}{Zhiwei Li}, \bibinfo{person}{Guodong Long}, {and} \bibinfo{person}{Tianyi Zhou}.} \bibinfo{year}{2024}\natexlab{}.
\newblock \showarticletitle{Federated Recommendation with Additive Personalization}. In \bibinfo{booktitle}{\emph{The Twelfth International Conference on Learning Representations}}.
\newblock


\bibitem[Li et~al\mbox{.}(2025a)]%
        {feddae}
\bibfield{author}{\bibinfo{person}{Zhiwei Li}, \bibinfo{person}{Guodong Long}, \bibinfo{person}{Tianyi Zhou}, \bibinfo{person}{Jing Jiang}, {and} \bibinfo{person}{Chengqi Zhang}.} \bibinfo{year}{2025}\natexlab{a}.
\newblock \showarticletitle{Personalized federated collaborative filtering: A variational autoencoder approach}. In \bibinfo{booktitle}{\emph{Proceedings of the AAAI Conference on Artificial Intelligence}}, Vol.~\bibinfo{volume}{39}. \bibinfo{pages}{18602--18610}.
\newblock
\href{https://doi.org/10.1609/aaai.v39i17.34047}{doi:\nolinkurl{10.1609/aaai.v39i17.34047}}


\bibitem[Lim et~al\mbox{.}(2025)]%
        {fcrec}
\bibfield{author}{\bibinfo{person}{Jaehyung Lim}, \bibinfo{person}{Wonbin Kweon}, \bibinfo{person}{Woojoo Kim}, \bibinfo{person}{Junyoung Kim}, \bibinfo{person}{Seongjin Choi}, \bibinfo{person}{Dongha Kim}, {and} \bibinfo{person}{Hwanjo Yu}.} \bibinfo{year}{2025}\natexlab{}.
\newblock \showarticletitle{Federated Continual Recommendation}. In \bibinfo{booktitle}{\emph{Proceedings of the 34th ACM International Conference on Information and Knowledge Management}}. \bibinfo{pages}{1798--1808}.
\newblock
\href{https://doi.org/10.1145/3746252.3761268}{doi:\nolinkurl{10.1145/3746252.3761268}}


\bibitem[McMahan et~al\mbox{.}(2017)]%
        {first_fl_paper}
\bibfield{author}{\bibinfo{person}{Brendan McMahan}, \bibinfo{person}{Eider Moore}, \bibinfo{person}{Daniel Ramage}, \bibinfo{person}{Seth Hampson}, {and} \bibinfo{person}{Blaise~Aguera y Arcas}.} \bibinfo{year}{2017}\natexlab{}.
\newblock \showarticletitle{Communication-efficient learning of deep networks from decentralized data}. In \bibinfo{booktitle}{\emph{Artificial intelligence and statistics}}. PMLR, \bibinfo{pages}{1273--1282}.
\newblock
\href{https://doi.org/10.48550/arXiv.1602.05629}{doi:\nolinkurl{10.48550/arXiv.1602.05629}}


\bibitem[Pardau(2018)]%
        {ccpa}
\bibfield{author}{\bibinfo{person}{Stuart~L Pardau}.} \bibinfo{year}{2018}\natexlab{}.
\newblock \showarticletitle{The california consumer privacy act: Towards a european-style privacy regime in the united states}.
\newblock \bibinfo{journal}{\emph{J. Tech. L. \& Pol'y}}  \bibinfo{volume}{23} (\bibinfo{year}{2018}), \bibinfo{pages}{68}.
\newblock


\bibitem[Paszke et~al\mbox{.}(2019)]%
        {pytorch}
\bibfield{author}{\bibinfo{person}{Adam Paszke}, \bibinfo{person}{Sam Gross}, \bibinfo{person}{Francisco Massa}, \bibinfo{person}{Adam Lerer}, \bibinfo{person}{James Bradbury}, \bibinfo{person}{Gregory Chanan}, \bibinfo{person}{Trevor Killeen}, \bibinfo{person}{Zeming Lin}, \bibinfo{person}{Natalia Gimelshein}, \bibinfo{person}{Luca Antiga}, {et~al\mbox{.}}} \bibinfo{year}{2019}\natexlab{}.
\newblock \showarticletitle{Pytorch: An imperative style, high-performance deep learning library}.
\newblock \bibinfo{journal}{\emph{Advances in neural information processing systems}}  \bibinfo{volume}{32} (\bibinfo{year}{2019}).
\newblock
\href{https://doi.org/10.5555/3454287.3455008}{doi:\nolinkurl{10.5555/3454287.3455008}}


\bibitem[Perifanis and Efraimidis(2022)]%
        {fedncf}
\bibfield{author}{\bibinfo{person}{Vasileios Perifanis} {and} \bibinfo{person}{Pavlos~S Efraimidis}.} \bibinfo{year}{2022}\natexlab{}.
\newblock \showarticletitle{Federated neural collaborative filtering}.
\newblock \bibinfo{journal}{\emph{Knowledge-Based Systems}}  \bibinfo{volume}{242} (\bibinfo{year}{2022}), \bibinfo{pages}{108441}.
\newblock
\href{https://doi.org/10.1016/j.knosys.2022.108441}{doi:\nolinkurl{10.1016/j.knosys.2022.108441}}


\bibitem[Voigt and Von~dem Bussche(2017)]%
        {gdpr}
\bibfield{author}{\bibinfo{person}{Paul Voigt} {and} \bibinfo{person}{Axel Von~dem Bussche}.} \bibinfo{year}{2017}\natexlab{}.
\newblock \showarticletitle{The eu general data protection regulation (gdpr)}.
\newblock \bibinfo{journal}{\emph{A practical guide, 1st ed., Cham: Springer International Publishing}} \bibinfo{volume}{10}, \bibinfo{number}{3152676} (\bibinfo{year}{2017}), \bibinfo{pages}{10--5555}.
\newblock
\href{https://doi.org/10.1007/978-3-319-57959-7}{doi:\nolinkurl{10.1007/978-3-319-57959-7}}


\bibitem[Volkovs et~al\mbox{.}(2017)]%
        {content_1}
\bibfield{author}{\bibinfo{person}{Maksims Volkovs}, \bibinfo{person}{Guang~Wei Yu}, {and} \bibinfo{person}{Tomi Poutanen}.} \bibinfo{year}{2017}\natexlab{}.
\newblock \showarticletitle{Content-based neighbor models for cold start in recommender systems}.
\newblock In \bibinfo{booktitle}{\emph{Proceedings of the recommender systems challenge 2017}}. \bibinfo{pages}{1--6}.
\newblock
\href{https://doi.org/10.1145/3124791.3124792}{doi:\nolinkurl{10.1145/3124791.3124792}}


\bibitem[Wahab et~al\mbox{.}(2022)]%
        {first_fedcold}
\bibfield{author}{\bibinfo{person}{Omar~Abdel Wahab}, \bibinfo{person}{Gaith Rjoub}, \bibinfo{person}{Jamal Bentahar}, {and} \bibinfo{person}{Robin Cohen}.} \bibinfo{year}{2022}\natexlab{}.
\newblock \showarticletitle{Federated against the cold: A trust-based federated learning approach to counter the cold start problem in recommendation systems}.
\newblock \bibinfo{journal}{\emph{Information Sciences}}  \bibinfo{volume}{601} (\bibinfo{year}{2022}), \bibinfo{pages}{189--206}.
\newblock
\href{https://doi.org/10.1016/j.ins.2022.04.027}{doi:\nolinkurl{10.1016/j.ins.2022.04.027}}


\bibitem[Wang and Blei(2011)]%
        {citeulike}
\bibfield{author}{\bibinfo{person}{Chong Wang} {and} \bibinfo{person}{David~M Blei}.} \bibinfo{year}{2011}\natexlab{}.
\newblock \showarticletitle{Collaborative topic modeling for recommending scientific articles}. In \bibinfo{booktitle}{\emph{Proceedings of the 17th ACM SIGKDD international conference on Knowledge discovery and data mining}}. \bibinfo{pages}{448--456}.
\newblock
\href{https://doi.org/10.1145/2020408.2020480}{doi:\nolinkurl{10.1145/2020408.2020480}}


\bibitem[Wang et~al\mbox{.}(2017)]%
        {dcn}
\bibfield{author}{\bibinfo{person}{Ruoxi Wang}, \bibinfo{person}{Bin Fu}, \bibinfo{person}{Gang Fu}, {and} \bibinfo{person}{Mingliang Wang}.} \bibinfo{year}{2017}\natexlab{}.
\newblock \showarticletitle{Deep \& cross network for ad click predictions}.
\newblock In \bibinfo{booktitle}{\emph{Proceedings of the ADKDD'17}}. \bibinfo{pages}{1--7}.
\newblock
\href{https://doi.org/10.1145/3124749.3124754}{doi:\nolinkurl{10.1145/3124749.3124754}}


\bibitem[Wei et~al\mbox{.}(2017)]%
        {cold_cf_1}
\bibfield{author}{\bibinfo{person}{Jian Wei}, \bibinfo{person}{Jianhua He}, \bibinfo{person}{Kai Chen}, \bibinfo{person}{Yi Zhou}, {and} \bibinfo{person}{Zuoyin Tang}.} \bibinfo{year}{2017}\natexlab{}.
\newblock \showarticletitle{Collaborative filtering and deep learning based recommendation system for cold start items}.
\newblock \bibinfo{journal}{\emph{Expert systems with applications}}  \bibinfo{volume}{69} (\bibinfo{year}{2017}), \bibinfo{pages}{29--39}.
\newblock
\href{https://doi.org/10.1016/j.eswa.2016.09.040}{doi:\nolinkurl{10.1016/j.eswa.2016.09.040}}


\bibitem[Wu et~al\mbox{.}(2022)]%
        {fedpergnn}
\bibfield{author}{\bibinfo{person}{Chuhan Wu}, \bibinfo{person}{Fangzhao Wu}, \bibinfo{person}{Lingjuan Lyu}, \bibinfo{person}{Tao Qi}, \bibinfo{person}{Yongfeng Huang}, {and} \bibinfo{person}{Xing Xie}.} \bibinfo{year}{2022}\natexlab{}.
\newblock \showarticletitle{A federated graph neural network framework for privacy-preserving personalization}.
\newblock \bibinfo{journal}{\emph{Nature Communications}} \bibinfo{volume}{13}, \bibinfo{number}{1} (\bibinfo{year}{2022}), \bibinfo{pages}{3091}.
\newblock
\href{https://doi.org/10.1038/s41467-022-30714-9}{doi:\nolinkurl{10.1038/s41467-022-30714-9}}


\bibitem[Zhang et~al\mbox{.}(2023)]%
        {pfedrec}
\bibfield{author}{\bibinfo{person}{Chunxu Zhang}, \bibinfo{person}{Guodong Long}, \bibinfo{person}{Tianyi Zhou}, \bibinfo{person}{Peng Yan}, \bibinfo{person}{Zijian Zhang}, \bibinfo{person}{Chengqi Zhang}, {and} \bibinfo{person}{Bo Yang}.} \bibinfo{year}{2023}\natexlab{}.
\newblock \showarticletitle{Dual personalization on federated recommendation}. In \bibinfo{booktitle}{\emph{Proceedings of the Thirty-Second International Joint Conference on Artificial Intelligence}}. \bibinfo{pages}{4558--4566}.
\newblock
\href{https://doi.org/10.24963/ijcai.2023/507}{doi:\nolinkurl{10.24963/ijcai.2023/507}}


\bibitem[Zhang et~al\mbox{.}(2024a)]%
        {gpfedrec}
\bibfield{author}{\bibinfo{person}{Chunxu Zhang}, \bibinfo{person}{Guodong Long}, \bibinfo{person}{Tianyi Zhou}, \bibinfo{person}{Zijian Zhang}, \bibinfo{person}{Peng Yan}, {and} \bibinfo{person}{Bo Yang}.} \bibinfo{year}{2024}\natexlab{a}.
\newblock \showarticletitle{Gpfedrec: Graph-guided personalization for federated recommendation}. In \bibinfo{booktitle}{\emph{Proceedings of the 30th ACM SIGKDD Conference on Knowledge Discovery and Data Mining}}. \bibinfo{pages}{4131--4142}.
\newblock
\href{https://doi.org/10.1145/3637528.3671702}{doi:\nolinkurl{10.1145/3637528.3671702}}


\bibitem[Zhang et~al\mbox{.}(2024b)]%
        {ifedrec}
\bibfield{author}{\bibinfo{person}{Chunxu Zhang}, \bibinfo{person}{Guodong Long}, \bibinfo{person}{Tianyi Zhou}, \bibinfo{person}{Zijian Zhang}, \bibinfo{person}{Peng Yan}, {and} \bibinfo{person}{Bo Yang}.} \bibinfo{year}{2024}\natexlab{b}.
\newblock \showarticletitle{When federated recommendation meets cold-start problem: Separating item attributes and user interactions}. In \bibinfo{booktitle}{\emph{Proceedings of the ACM Web Conference 2024}}. \bibinfo{pages}{3632--3642}.
\newblock
\href{https://doi.org/10.1145/3589334.3645525}{doi:\nolinkurl{10.1145/3589334.3645525}}


\bibitem[Zhang et~al\mbox{.}(2019)]%
        {rs_survey}
\bibfield{author}{\bibinfo{person}{Shuai Zhang}, \bibinfo{person}{Lina Yao}, \bibinfo{person}{Aixin Sun}, {and} \bibinfo{person}{Yi Tay}.} \bibinfo{year}{2019}\natexlab{}.
\newblock \showarticletitle{Deep learning based recommender system: A survey and new perspectives}.
\newblock \bibinfo{journal}{\emph{ACM computing surveys (CSUR)}} \bibinfo{volume}{52}, \bibinfo{number}{1} (\bibinfo{year}{2019}), \bibinfo{pages}{1--38}.
\newblock
\href{https://doi.org/10.1145/3285029}{doi:\nolinkurl{10.1145/3285029}}


\bibitem[Zhou et~al\mbox{.}(2025)]%
        {fediar}
\bibfield{author}{\bibinfo{person}{Pengyang Zhou}, \bibinfo{person}{Chaochao Chen}, \bibinfo{person}{Weiming Liu}, \bibinfo{person}{Wenkai Shen}, \bibinfo{person}{Xinting Liao}, \bibinfo{person}{Huarong Deng}, \bibinfo{person}{Zhihui Fu}, \bibinfo{person}{Jun Wang}, \bibinfo{person}{Wu Wen}, {and} \bibinfo{person}{Xiaolin Zheng}.} \bibinfo{year}{2025}\natexlab{}.
\newblock \showarticletitle{Joint item embedding dual-view exploration and adaptive local-global fusion for federated recommendation}. In \bibinfo{booktitle}{\emph{Proceedings of the 48th International ACM SIGIR Conference on Research and Development in Information Retrieval}}. \bibinfo{pages}{424--434}.
\newblock
\href{https://doi.org/10.1145/3726302.3730016}{doi:\nolinkurl{10.1145/3726302.3730016}}


\bibitem[Zhou et~al\mbox{.}(2023)]%
        {cold_cf_2}
\bibfield{author}{\bibinfo{person}{Zhihui Zhou}, \bibinfo{person}{Lilin Zhang}, {and} \bibinfo{person}{Ning Yang}.} \bibinfo{year}{2023}\natexlab{}.
\newblock \showarticletitle{Contrastive collaborative filtering for cold-start item recommendation}. In \bibinfo{booktitle}{\emph{Proceedings of the ACM web conference 2023}}. \bibinfo{pages}{928--937}.
\newblock
\href{https://doi.org/10.1145/3543507.3583286}{doi:\nolinkurl{10.1145/3543507.3583286}}


\bibitem[Zhu et~al\mbox{.}(2020)]%
        {heater}
\bibfield{author}{\bibinfo{person}{Ziwei Zhu}, \bibinfo{person}{Shahin Sefati}, \bibinfo{person}{Parsa Saadatpanah}, {and} \bibinfo{person}{James Caverlee}.} \bibinfo{year}{2020}\natexlab{}.
\newblock \showarticletitle{Recommendation for new users and new items via randomized training and mixture-of-experts transformation}. In \bibinfo{booktitle}{\emph{Proceedings of the 43rd International ACM SIGIR conference on research and development in Information Retrieval}}. \bibinfo{pages}{1121--1130}.
\newblock
\href{https://doi.org/10.1145/3397271.3401178}{doi:\nolinkurl{10.1145/3397271.3401178}}


\end{thebibliography}

\appendix

\section{Why Clipping + Gaussian Noise Satisfies $(\varepsilon,\delta)$-LDP}
\label{appendix:ldp}

A randomized mechanism $\mathcal{M}$ satisfies $(\varepsilon,\delta)$-local differential privacy (LDP) if for any two inputs
$\psi,\psi'$ and any measurable set $\mathcal{H}$,
\begin{align}
\label{eq:definition_ldp}
\Pr[\mathcal{M}(\psi)\in\mathcal{H}]
\;\le\;
e^{\varepsilon}\Pr[\mathcal{M}(\psi')\in\mathcal{H}] + \delta .
\end{align}
In our local update, the client computes the gradient
$G=\nabla_{Q_u}\mathcal{L}_u \in \mathbb{R}^{|\mathcal I^W|\times d}$
and releases a privatized gradient via clipping and Gaussian noise.
Define the clipping multiplier and the clipped gradient as
\begin{align}
\alpha(G)\triangleq \min\!\left(1,\frac{C}{\|G\|_F}\right),\qquad
\bar G \triangleq \alpha(G)\,G,
\end{align}
so that $\|\bar G\|_F\le C$. The released (private) gradient is
\begin{align}
\tilde G \;=\; \bar G + Z,
\qquad
\mathrm{vec}(Z) \sim \mathcal{N}\!\left(0, s^2 I_{|\mathcal I^W|d}\right),
\qquad
s=\sigma C,
\end{align}
where $\mathrm{vec}(\cdot)$ denotes the vectorization operator.

\noindent \textbf{Step 1: Sensitivity bound induced by clipping.}
Let $f(G)=\bar G$ denote the clipped mapping.
Since $\|f(G)\|_F \le C$ for any $G$, for any two inputs $G,G'$ we have
\begin{align}
\|f(G)-f(G')\|_F \;\le\; \|f(G)\|_F + \|f(G')\|_F \;\le\; 2C.
\end{align}
Thus, the $\ell_2$-sensitivity (after vectorization) is bounded by
\begin{align}
\Delta_2 \;\triangleq\; \sup_{G,G'} \|f(G)-f(G')\|_F \;\le\; 2C.
\label{eq:app_sensitivity}
\end{align}


\noindent \textbf{Step 2: Gaussian mechanism calibration.}
Consider the Gaussian mechanism $\mathcal{M}(G)=f(G)+Z$ with $\mathrm{vec}(Z)\sim\mathcal{N}\!(0,s^2 I_{|\mathcal I^W|d})$.
By the standard Gaussian mechanism theorem (e.g., \cite{ldp_125}), for any $\delta\in(0,1)$, if $f$ has $\ell_2$-sensitivity
at most $\Delta_2$, then adding Gaussian noise with standard deviation
\begin{align}
s \;\ge\; \frac{\Delta_2}{\varepsilon}\sqrt{2\log\!\left(\frac{1.25}{\delta}\right)}
\label{eq:app_gauss_calib}
\end{align}
is sufficient to satisfy the privacy condition in Eq.~\eqref{eq:definition_ldp} when the mechanism is applied locally to a single client
(i.e., to that client’s private data through its local gradient).
Hence, each local step provides an $(\varepsilon,\delta)$-LDP guarantee.

\noindent \textbf{Instantiation for clipped gradients.}
Combining Eq.~\eqref{eq:app_sensitivity} and Eq.~\eqref{eq:app_gauss_calib} with $\Delta_2\le 2C$ and $s=\sigma C$ yields
\begin{align}
\sigma \;\ge\; \frac{2}{\varepsilon}\sqrt{2\log\!\left(\frac{1.25}{\delta}\right)}.
\label{eq:app_sigma}
\end{align}
Therefore, releasing $\tilde G=\bar G+Z$ with $\bar G=G\cdot \min(1,C/\|G\|_F)$ and
$\mathrm{vec}(Z)\sim\mathcal{N}(0,\sigma^2 C^2 I_{|\mathcal I^{W}|d})$ satisfies $(\varepsilon,\delta)$-LDP for each local step.
Finally, the server observes $\Delta Q_u=-\eta \tilde G$ (equivalently $Q_u^{\mathrm{new}}=Q_u-\eta\tilde G$),
which is deterministic post-processing of $\tilde G$; hence it preserves the same $(\varepsilon,\delta)$-LDP guarantee.

In summary, clipping enforces $\|f(G)\|_F\le C$ and thus $\Delta_2 \le 2C$ (Eq.~\eqref{eq:app_sensitivity}),
while Eq.~\eqref{eq:app_gauss_calib} calibrates the Gaussian noise accordingly to satisfy Eq.~\eqref{eq:definition_ldp}.
A detailed proof of the Gaussian calibration (including the constant $1.25$) can be found in \cite{ldp_125}.

\begin{algorithm}[h]
\small
\caption{PMFRec: Warm Training Phase}
\label{algo:warm}
    \begin{algorithmic}[1]
        \Statex \textbf{ServerExecute}:
        \State Initialize a multi-view encoder and a personalized encoder.
        \For{$t=1, 2, ..., T$}
            \State Sample participating users $\mathcal U^t$ from $\mathcal U$
            \State Compute global warm-item representations $P_g$ with Eq.~\eqref{eq:FM_def}.
        
            \ForAll{$u\in \mathcal U^t$ \textbf{in parallel}}
                \If{$t=1$}
                    \State Compute $Z_u$ as $P_g$.
                \Else
                    \State Compute $Z_u$ with Eq.~\eqref{eq:Zu}
                \EndIf
                
                \State  $Q_u^{\mathrm{new}} \leftarrow \mathrm{ClientUpdate}(Z_u,t)$ 

                \State Update parameters of $h_u$ with Eq.~\eqref{eq:l_p}
                \State Do cumulative aggregation with Eq.~\eqref{eq:cum_agg}
                
            \EndFor
        
            \For{$e=1,...,E_{\mathrm{server}}$}
                \State Compute $\mathcal L_{M}$ with Eq.~\eqref{eq:LM}, $\mathcal L_{\mathrm{ortho}}$ with Eq.~\eqref{eq:L_ortho}, and $\mathcal L_{\mathrm{lb}}$ with Eq.~\eqref{eq:L_lb}
                \State Update parameters of $F_M$ with Eq.~\eqref{eq:server_total_loss}
            \EndFor
        \EndFor
    \end{algorithmic}

    \begin{algorithmic}[1]
        \Statex \textbf{ClientUpdate}($Z_u$, t):
        \If {$t=1$}
            \State Initialize user embedding $e_u$ and scoring function $\Theta_u$
        \Else
            \State Initialize user embedding and scoring function with the latest updates
        \EndIf
        
        \State Initialize $Q_u$ with $Z_u$

        \State Sample negative feedback $\mathcal{D}_u^{(W,-)}$ from $\mathcal{I}^W$
        \State $\mathcal{B}\leftarrow$ (split \{$\mathcal{D}_u^{(W,+)}\cup\mathcal{D}_u^{(W,-)}\}$ into batches)

        \For {$e$ from $1$ to $E_{\mathrm{local}}$}
            \For {batch $b \in \mathcal{B}$}
                \State Compute $\mathcal L_u$ with Eq.~\eqref{eq:local_loss}
                \State Update ($Q_u, e_u,\Theta_u$) with Eq.~\eqref{eq:client_updates}
            \EndFor
        \EndFor
        \State \textbf{return} $Q_u^{\mathrm{new}}$ to server
    \end{algorithmic}
\end{algorithm}

\end{document}